\documentclass[11pt,a4paper]{article}
\usepackage[utf8]{inputenc}
\def\parfrac#1#2{{\left(\frac{#1}{#2}\right) }} 
\usepackage{caption}
\usepackage{subcaption}
\usepackage{jcappub}
\usepackage{bm}
\usepackage{epsfig}
\usepackage{bbold}
\usepackage{graphicx}
\usepackage{amsmath}
\usepackage{amssymb}
\usepackage{amsbsy}
\usepackage{color}
\usepackage{slashed}
\usepackage{afterpage}
\usepackage{psfrag}
\usepackage[noabbrev]{cleveref}
\usepackage{dsfont}
\usepackage{enumerate}
\usepackage[shortlabels]{enumitem}
\usepackage[utf8]{inputenc}
\usepackage{amsmath}
\usepackage[dvipsnames]{xcolor}
\usepackage{graphicx}
\usepackage{braket}
\usepackage{physics}
\usepackage{siunitx}
\usepackage{float}
\usepackage{tikz-feynman}
\tikzset{snake/.style={decorate, decoration=snake}}
\usepackage{tikz}
\usepackage{placeins}
\usepackage{comment}
\usepackage{tikz-3dplot}
\usetikzlibrary{arrows.meta,calc,3d,shadings,positioning,decorations.text}
\newcommand{\Mpl}{M_{\rm Pl}}
\newcommand{\mpl}{\Mpl}

\newcommand{\rhoa}{\rho_a}
\newcommand{\rhoc}{\rho_c}
\newcommand{\fa}{f_a}
\newcommand{\fdm}{f_{\rm DM}}

\title{
\begin{center}
Stimulated capture of axion dark matter around primordial black holes
\end{center}
}

\author[a]{Nuno P.~Branco}
\author[a]{Jo\~ao G.~Rosa}

\affiliation[a]{Univ Coimbra, Faculdade de Ci\^encias e Tecnologia da Universidade de Coimbra and CFisUC, Rua Larga, 3004-516 Coimbra, Portugal}

\emailAdd{nuno.branco@student.fis.uc.pt}
\emailAdd{jgrosa@uc.pt}

\makeatletter
\gdef\@fpheader{}
\makeatother

\abstract{
We study the stimulated capture of axion and axion-like particles around primordial black holes, mediated by quartic self-interactions. We identify  the parametric regions where Bose-enhanced growth overcomes cosmological dilution and derive analytic criteria for efficient formation and long-term survival of the hydrogenic ground-state cloud, confronting them with current bounds on axion--photon couplings.  We discuss observational implications of these PBH$+$cloud systems, including resonant axion-to-photon conversion in neutron star magnetospheres and finite-size microlensing distortions. In the former case, the resonance condition selects a conversion shell outside the neutron star surface, and we show that the large axion occupation number of the cloud can compensate for the small conversion probability. The resulting transient line flux densities can exceed nominal VLA and ALMA sensitivities in regions where the Galactic encounter rate is observationally relevant. Independently, when the cloud size becomes comparable to or larger than the Einstein radius, finite-size effects can modify the standard point-lens microlensing light curve.
}

\begin{document}
\maketitle

% =========================================================
\section{Introduction}\label{sec:introduction}
% =========================================================

Despite overwhelming gravitational evidence for dark matter (DM), its microscopic nature remains unknown.
A broad and well-motivated possibility is that DM is a light bosonic field: for sufficiently small masses,
the phase-space occupation number in astrophysical environments is very large, so DM behaves as a coherent
classical wave \cite{Hui:2021tkt}. Axions and axion-like particles (ALPs) provide concrete realizations of this scenario.
In the non-relativistic regime, the dynamics is accurately captured by a Schr\"odinger--Poisson or Gross--Pitaevskii-type
description, including self-interactions. These self-interactions are generic in axion/ALP
theories and can become dynamically relevant in overdense regions, enabling the formation of DM substructure on
scales much smaller than the host halo.

A particularly interesting class of bosonic overdensities are ``gravitational atoms'': bound configurations supported
primarily by an external gravitational potential, in direct analogy with hydrogenic bound states in a Coulomb field.
Around rotating (Kerr) black holes, the best-known mechanism to populate such levels is superradiance, which
exponentially amplifies modes satisfying $\omega < m \Omega_H$ (with $m$ the azimuthal quantum number and $\Omega_H$
the horizon angular velocity) by extracting energy and angular momentum from the black hole. Since the gravitational
potential is Coulomb-like at large radii, the quasi-bound spectrum for a (spinless) field of mass $\mu$ around a black hole
of mass $M$ takes a Hydrogen-like form in the non-relativistic limit, $\hbar\omega_n \simeq \mu c^2\left(1-\alpha^2/2n^2\right)$ where $n$ is the principal quantum number and,
\begin{equation}
\alpha \equiv \frac{\mu G M_{\rm{PBH}}}{\hbar c}
\simeq 4\times 10^{-10}\,
\parfrac{M_{\rm{PBH}}}{10^{14}\,\mathrm{kg}}\,
\parfrac{\mu}{10^{-3}\,\mathrm{eV}} \, .
\label{eq:coupling}
\end{equation}
This has attracted a significant interest in the recent literature as a novel laboratory to probe
both Standard Model and beyond-the-Standard-Model particles, including in particular the QCD axion and other ALPs
\cite{Arvanitaki:2009fg,Arvanitaki:2010sy,Rosa:2012uz,Witek:2012tr,Brito:2014wla,Arvanitaki:2016qwi,
Brito:2017wnc,Rosa:2017ury,Baumann:2018vus,Ikeda:2018nhb,Berti:2019wnn,Baumann:2019eav,Cardoso:2018tly,
Edwards:2019tzf,Mehta:2020kwu,Blas:2020nbs,Tong:2022bbl,Branco:2023frw,Neves:2024rju}.

Superradiance, however, is not the only route to form gravitational atoms. Recent work identified a generic and efficient self-interaction-mediated formation mechanism in which quartic self-interactions
induce $2\to 2$ scattering among DM waves in an external potential, allowing one of the scattered quanta to become bound
while the other carries away the excess energy \cite{Budker:2023sex}. The bound mass $M_{\rm cloud}$ then evolves schematically as,
\begin{equation} \label{mass_evolution}
\dot{M}_{\rm cloud} = C + (\Gamma_1-\Gamma_2)\,M_{\rm cloud} \, ,
\end{equation}
where $C>0$ accounts for direct capture, while $\Gamma_1$ and $\Gamma_2$ encode, respectively, stimulated capture
(Bose enhancement) and the inverse ``stripping'' process\footnote{Although we follow the nomenclature used in \cite{Budker:2023sex}, we note that this is equivalent to collisional ionization with ambient DM axions.}. The sign of the net rate $\Gamma\equiv \Gamma_1-\Gamma_2$
is controlled by a single focusing parameter,
\begin{equation}
\xi \equiv \frac{2\pi}{kR_\star} \simeq 2\pi\,\frac{\alpha}{v_{\rm DM}} \, ,
\end{equation}
where $v_{\rm DM}$ denotes the asymptotic DM velocity in the BH reference frame, which we assume to coincide with the cosmological frame (where the axion field is homogeneous)\footnote{In other words, we are assuming that PBHs are formed at rest in this frame, with negligible natal ``kicks''.}. In the above expression, $k$ is the momentum of the wave and $R_\star$ is the gravitational ``Bohr radius'' (for the ground state $R_\star = 1/(\mu\alpha)$). For $\xi\lesssim 1$, $\Gamma_2> \Gamma_1$, implying that the cloud growth is limited and leads at most
to a dilute bound component. In contrast, for $\xi\gtrsim 1$ gravitational focusing enhances the incoming waves within
$r\sim R_\star$, suppresses stripping relative to stimulated capture, and can trigger exponential growth of the bound
population. In this stimulated regime, the cloud is dominated by the spherically symmetric ground state ($1s$-like),
with excited states typically subdominant and/or cascading efficiently to the ground state \cite{Budker:2023sex}.
For the ground state, with quantum numbers $(n,l,m)=(1,0,0)$ in the atomic language, the stimulated growth rate is given by,
\begin{equation}
\Gamma_1\simeq \frac{3}{64\,\fa^{4}}\,
\frac{\rho_{a}^{\,2}}{\mu^{3}\,v_{a}^{2}} \, ,
\label{eq: stimulatedrate}
\end{equation}
where $f_a$ denotes the axion-like decay constant, which parametrizes the strength of the self-interactions, and $\rho_a$ the ambient axion/ALP DM density.

In this work we extend the stimulated capture mechanism from stellar potentials to (primordial) black holes (PBHs). 
Extended clouds around black holes can induce small but measurable departures from Keplerian motion of nearby orbiting stars, 
making wide black hole binaries discovered by Gaia a promising probe of these structures \cite{Branco:2025znp}.
We consider non-rotating (Schwarzschild) black holes as external gravitational sources and study the conditions under which an ambient axion/ALP population can build up a hydrogenic cloud via self-interaction-mediated capture. The PBH case is particularly well motivated given that they experience axion DM conditions in the early universe, namely much larger densities and smaller velocity dispersion than in late-forming halos, assuming a dominant misalignment axion DM production mechanism. Both effects can significantly enhance the stimulated capture rate in Eq.~(\ref{eq: stimulatedrate}). In addition, the cosmological redshift of the axion velocities implies a dynamical evolution between the $\xi\lesssim1$ and $\xi\gtrsim 1$ regimes, with stimulated growth naturally becoming a relevant capture mechanism.
We focus on the physically relevant situation in which axions (either ALPs or the QCD axion  whose mass is temperature-dependent prior to the QCD crossover) constitute (most of) the DM abundance, and we determine when stimulated
growth becomes efficient compared to Hubble expansion.

The formation of dense axion clouds around PBHs has direct phenomenological implications. First, it can enhance radiative signatures associated with axion couplings, including narrow spectral features from axion conversion into photons in regions of large overdensity. Second, the cloud can contribute nontrivially to gravitational lensing: because the cloud is extended on a scale set by $R_\star$, PBH+cloud systems can depart from the point-lens approximation and yield characteristic
microlensing distortions when the cloud size becomes comparable to the Einstein radius. 

This paper is organized as follows. In the next section we introduce the axion/ALP framework and summarize the stimulated capture rates relevant for the lowest bound state. We then determine the cosmological conditions for efficient stimulated capture and long-term cloud survival. In
Section \ref{sec:Detection} we discuss potential observational signatures of PBH-seed axion clouds, focusing on axion-to-photon conversion in neutron star magnetospheres and gravitational microlensing.  We summarize our results and discuss future prospects in Section \ref{sec:conclusions}. In appendix \ref{app:dense_growth}, we describe in more technical detail the formation and evolution of axion clouds. We work in natural units $\hbar=c=1$ and use the %(non-reduced) 
Planck mass $G = M_{\rm {Pl}}^{-2}$ throughout.

% =========================================================
\section{PBH-axion cloud formation and evolution}\label{sec:axions_general}
% =========================================================

In this section we study light bosonic DM candidates in the axion family, characterized by a (possibly temperature-dependent)
mass scale $\mu$ and a decay constant $f_a$ controlling the leading self-interactions. We consider both generic ALPs and the QCD axion. We assume the axion(-like) field constitutes a fraction $\fdm$ of the total DM abundance,
\begin{equation}
\rhoa = \fdm\,\rho_{\rm DM} \, ,
\label{eq:rhoa_fraction}
\end{equation}
alongside a sub-dominant population of PBHs that can act as seeds for gravitationally bound axion clouds, determining under which conditions their mass can grow efficiently through stimulated, self-interaction-driven capture.

According to Eq.~(\ref{mass_evolution}), axion clouds start growing through direct capture. The capture rate is, in fact, constant despite the cosmological redshift as we show in the appendix, which implies a linear growth of the axion cloud's mass. Since $\Gamma<0$, this would eventually be balanced by stripping processes if the system remained in the unfocused regime $\xi\lesssim 1$. However, since the velocity of the cosmological axions decreases as the universe expands, eventually gravitational focusing becomes important, $\xi>1$, making stimulated capture dominate over stripping processes and leading, hence, to an exponential growth of the cloud. While the realistic transition between the two capture regimes is smooth, it is, for our purposes, sufficient to model it as a hard transition at the instant $t_*$ at which $\xi=1$:
\begin{equation}
v_a(t_*) = 2\pi\,\alpha\,,
\label{eq:threshold_master}
\end{equation}
Since $\Gamma_{1}\propto a^{-4}$ for non-relativistic axions, exponential growth is effectively limited to timescales of order $t_*$, so that stimulated capture is only efficient if $\Gamma_1(t_*)\gtrsim H(t_*)$.

The axion cloud mass cannot, however, reach arbitrarily large values. Once the axion density reaches a critical value $\rho_{\rm cloud}\simeq 16\alpha^2\mu^2f_a^2$, the self-interaction energy becomes comparable to the gravitational binding energy, and their attractive nature makes the cloud first collapse and then undergo a {\it bosenova} explosion, where a fraction of the cloud mass is released through the emission of relativistic axions \cite{Arvanitaki:2009fg, Yoshino:2012kn, Yoshino:2015nsa, Omiya:2020vji,Omiya:2022gwu, Omiya:2022mwv}. After the bosenova, the cloud density becomes sub-critical and growth may resume until a new bosenova occurs. This process may, in principle, repeat multiple times on time scales of order $t_*$, although the dilution of the cosmological axion density eventually quenches any further growth and bosenova explosions. The main conclusion is that the cloud mass saturates close a maximum value
\begin{equation}
M_{\rm cloud}^{\rm max}\simeq {8\pi\over \alpha} {f_a^2 \over \mu}\simeq {8\pi\over \alpha^2}\left({f_a\over M_P}\right)^2 M_{\rm PBH}\, ,
\label{eq: maxmass}
\end{equation}
which may therefore correspond to a parametrically large fraction of the PBH mass.

In appendix \ref{app:dense_growth}, we give a detailed description of the cloud evolution in both the radiation and matter eras, showing that stimulated capture needs to be efficient over 40-70 e-folds for an axion cloud to reach its maximum mass, depending on the parametric range. For simplicity, we impose a global condition for the full growth of an axion cloud\footnote{We note that fixing $60$ e-folds across the full parameter or imposing the exact number e-folds required to reach the maximum cloud mass does not significantly change our results.}:
\begin{equation}
\left.{\Gamma_1\over H}\right|_{t_*}\gtrsim 60\beta^{-1}\, .
\label{eq:growth_criterion}
\end{equation}
with $\beta=0.5$ (0.4) if stimulated capture becomes relevant in the radiation (matter) era.

We are mostly interested in axion clouds that survive until the present day, with potential observational signatures. PBH-axion clouds may be depleted in two main ways. First, when the first DM halos start collapsing gravitationally at late times, the axion velocity dispersion increases, such that eventually $\xi<1$. This implies that stripping processes become dominant over stimulated capture ($\Gamma_1<\Gamma_2$ in Eq.~(\ref{mass_evolution})). For $\xi\ll 1$, we have $\Gamma_2\simeq 2\Gamma_1$, so that the cloud is collisionally depleted on a timescale given by:
\begin{equation}
-\frac{1}{\Gamma}
\simeq
9\,\mathrm{Gyr}
\left(
\frac{f_a}{10^8\,\mathrm{GeV}}
\right)^4
\left(
\frac{10^{-12}\,\mathrm{eV}}{\mu}
\right)
\left(
\frac{M_\odot}{M_{\rm{PBH}}}
\right)^4
\left(
\frac{0.4\,\mathrm{GeV}/\mathrm{cm}^3}{\rho_{\mathrm{a}}}
\right)^2
\left(
\frac{v_{\mathrm{a}}}{240\,\mathrm{km}/\mathrm{s}}
\right)^6~.
\end{equation}

Second, axion bound states around a BH are intrinsically unstable, given the ingoing boundary conditions at the horizon, which in fact distinguishes the BH and stellar potential cases. This is manifest in the negative imaginary part of the quasi-bound state frequency spectrum, which corresponds to the reabsorption of the axion cloud by the black hole\footnote{For spinning Kerr BHs, $\Gamma_{\rm{abs}}\propto \omega-m\Omega$, so that $\Gamma_{\rm{abs}}<0$ for bound states with $\omega<m\Omega$, which corresponds to superradiant amplification. For Schwarzschild BHs, on which we focus our analysis, $\Gamma_{\rm{abs}}>0$ for all states. }. For the ground state, we have in the non-relativistic regime $\alpha\ll1$ \cite{Detweiler:1980uk}:
\begin{equation}
\Gamma_{\rm abs}^{-1}\simeq 4\times10^4
\,{\rm yrs}\,
\left(\frac{10^{-13}\,{\rm eV}}{\mu}\right)^6
\parfrac{M_\odot}{M_{\rm{PBH}}}^5 ~.
\label{eq:absorptionrate}
\end{equation}
To ensure the survival of axion clouds until the present day, we therefore require that the late stripping and reabsorption timescales exceed the age of the universe, in the former case taking a value of the axion density and velocity comparable to the corresponding DM values in the solar neighborhood.

% =========================================================
\subsection{Axion-like particles}\label{subsec:ALPs}
% =========================================================

We first consider ALPs with a constant mass $\mu$  throughout the cosmic history. In this section we assume the ALP population is predominantly produced by misalignment, so that after the onset of oscillations the velocity dispersion redshifts from an initially order-one value to a very small value at later times but before structure formation). The relevant cosmological inputs are the ALP density $\rhoa(a)$ and velocity dispersion $v_a(a)$ after the onset of oscillations. The critical density today is $\rhoc^{0}=3H_0^{2}\mpl^{2}/(8\pi)$, and the present DM density is
$\rho_{\rm DM,0}=\Omega_{DM,0}\,\rho_{c,0}$. Since ALPs are non-relativistic after oscillations begin, $\rhoa\propto a^{-3}$,
\begin{equation}
\rhoa(a)=\fdm\,\rho_{\rm DM}(a)
=\frac{3\,\fdm\,\Omega_{DM,0}\,H_0^{2}\mpl^{2}}{8\pi}\,a^{-3}\, .
\label{eq:rhoa_of_a_compact}
\end{equation}
In virialized halos one expects $v_a\simeq 10^{-3}$, whereas in the early Universe the velocity dispersion can be much smaller, enhancing $\Gamma_{1}$ through the $1/v_a^{2}$ factor.
After the field starts oscillating at $H(T_{\rm osc})\simeq\mu$ during radiation domination,
\begin{equation}
T_{\rm osc}\simeq 2.7\times 10^3\,{\rm GeV}\left(\frac{\mu}{10^{-3}\,{\rm eV}}\right)^{1/2}
g_{*}^{-1/4}(T_{\rm osc})\,,
\label{eq:Tosc_compact}
\end{equation}
the typical physical momentum redshifts as $p\propto a^{-1}$, implying $v_a\propto a^{-1}$. Using entropy conservation,
$g_{*s}^{1/3}(T)\,T\,a=\mathrm{const.}$, one finds the convenient parametrization,
\begin{equation}
v_a(a)\simeq
1.4\times 10^{-16}\left(\frac{10^{-3} \rm{eV}}{\mu}\right)^{1/2}\frac{g_{*}^{1/4}(T_{\rm osc})}{a\,g_{*s}^{1/3}(T_{\rm osc})}\, .
\label{eq:v_of_a_compact}
\end{equation}
Here $g_*(T)$ and $g_{*s}(T)$ denote the number of  relativistic degrees of freedom in the
energy and entropy densities, respectively. Stimulated capture becomes significant when Eq.~\eqref{eq:threshold_master} is satisfied. Solving
~\eqref{eq:threshold_master} with~\eqref{eq:v_of_a_compact} and using $T(a)=T_{0}a^{-1}(g_{*s}^{0}/g_{*s})^{1/3}$ gives
\begin{equation}
T_* \simeq 7 \,{\rm keV}
\parfrac{M_{\rm{PBH}}}{10^{14}\,\mathrm{kg}}
\left(\frac{\mu}{10^{-3}\,{\rm eV}}\right)^{3/2}
\left(\frac{g_{*s}(T_{\rm osc})}{g_{*s}(T_*)}\right)^{1/3}g_{*}^{-1/4}(T_{\rm osc}) \, .
\label{eq:astar}
\end{equation}
For our purposes, the relevant quantity is the Hubble rate at the stimulated capture threshold, which depends on whether $t_*$ lies in radiation
or matter domination. Using $H\simeq 1/(2t)$ in radiation and the standard matching across matter-radiation equality at $t_{\rm{eq}}$, one obtains,
\begin{equation}
t_* \simeq
\begin{cases}
5.2\times 10^4\,{\rm s}\,
\parfrac{10^{14}\,\mathrm{kg}}{M_{\rm{PBH}}}^2
\left(\dfrac{10^{-3}\,{\rm eV}}{\mu}\right)^3\,,
& t<t_{\rm{eq}}\,,\\[6pt] 3.3\times 10^6\,{\rm yrs}\,
\parfrac{10^{24}\,\mathrm{kg}}{M_{\rm{PBH}}}^{3/2}
\left(\dfrac{ 10^{-13}\,{\rm eV}}{\mu}\right)^{9/4}\,,
& t>t_{\rm{eq}}\, .
\end{cases}
\label{eq:tstar_cases}
\end{equation}
We now evaluate the growth criterion \eqref{eq:growth_criterion} at $t_*$, which yields:
\begin{equation}
\left.\frac{\Gamma_{1}}{H}\right|_{t_*}
\simeq
\begin{cases}
6.3\times 10^2 \, \fdm^{2}\,
\parfrac{M_{\rm{PBH}}}{10^{14}\,\mathrm{kg}}^2
\left(\dfrac{10^{7}\,{\rm GeV}}{\fa}\right)^{4}
\left(\dfrac{\mu}{10^{-3}\,{\rm eV}}\right)\,
\dfrac{g_{*s}^{4/3}(T_{\rm osc})}{g_*(T_{\rm osc})}\,,
& t<t_{\rm{eq}}\,,\\[8pt]
1.4\times 10^2\,\fdm^{2}\,
\parfrac{M_{\rm{PBH}}}{10^{24}\,\mathrm{kg}}^{5/2}
\left(\dfrac{10^{9}\,{\rm GeV}}{\fa}\right)^{4}
\left(\dfrac{\mu}{10^{-13}\,{\rm eV}}\right)^{7/4}\,
\dfrac{g_{*s}^{3/2}(T_{\rm osc})}{g_*^{9/8}(T_{\rm osc})}\,,
& t>t_{\rm{eq}}\,.
\end{cases}
\label{eq:Gamma_over_H_cases}
\end{equation}

In Figure \ref{fig:alps_region} we show the regions in the $(\mu,f_a)$ plane (and the related axion-photon coupling, see below) for which ALP clouds can efficiently grow around PBHs in the sub-solar mass range. In addition to the growth criterion, we require that the threshold for stimulated capture is reached before the onset of structure formation at $z\simeq 30$ \cite{Klessen:2023qmc}, after which our analysis considering an oscillating axion condensate ceases to be valid.

\begin{figure}[t]
    \centering
\includegraphics[width=0.95\textwidth]{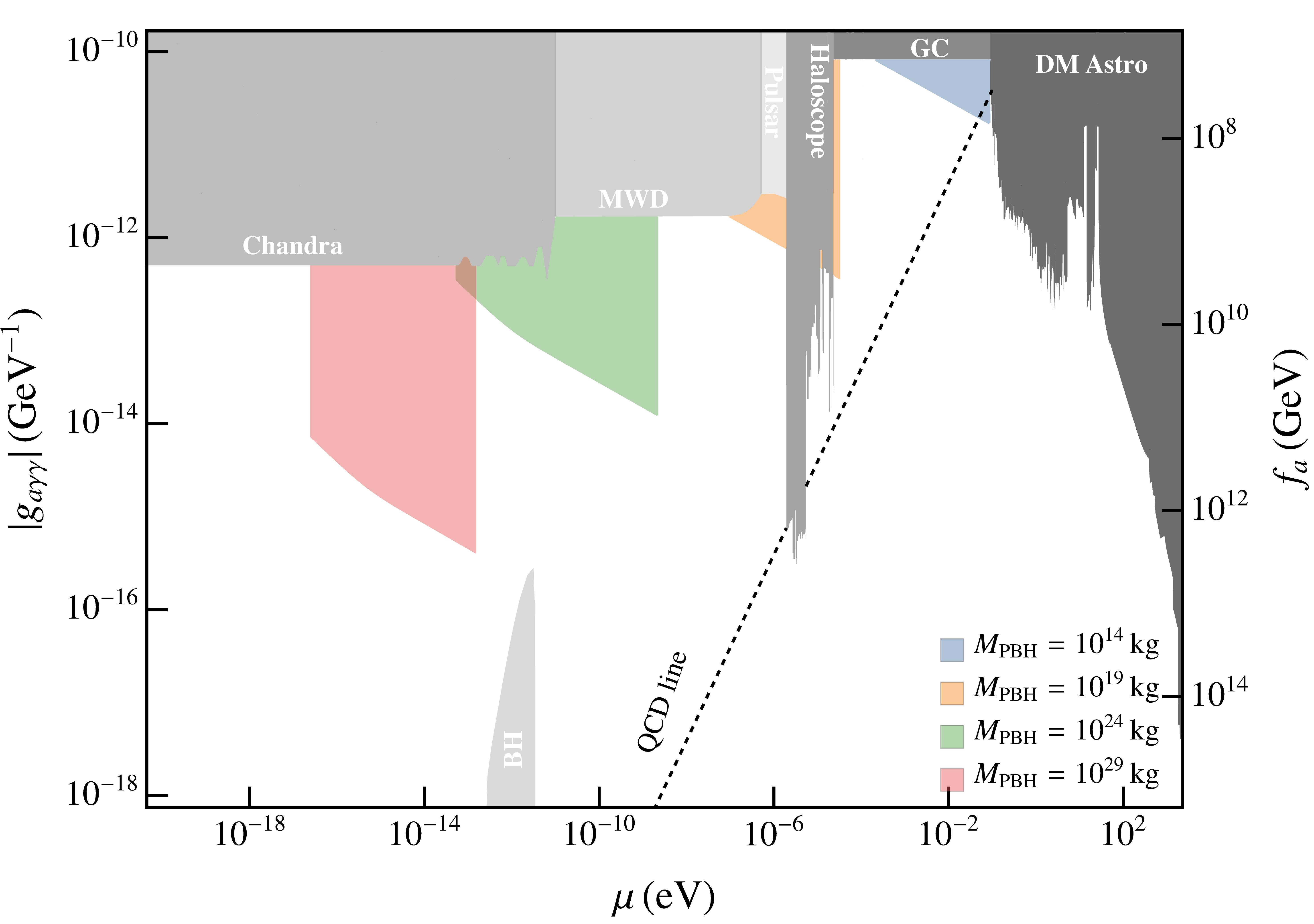}
    \caption{\small
Allowed regions for PBH-seeded stimulated growth of ALP clouds in the $(\mu,\lvert g_{a\gamma\gamma}\rvert)$ plane ($c_\gamma=1$). Colored patches correspond to different reference PBH masses.
The gray shaded areas summarize existing constraints on $|g_{a\gamma\gamma}|$ from astrophysical and laboratory searches, including DM/astrophysical probes, globular clusters, haloscopes, X-ray observations, magnetic white dwarfs, pulsars, and black hole superradiance \cite{AxionLimits,Gorghetto:2022ikz}.
The black dashed curve shows the QCD-axion line for reference.}
\label{fig:alps_region}
\end{figure}

For each value of the PBH mass, the viable parametric regions are characterized by a maximum axion mass, above which clouds are reabsorbed by the PBH before the present day, as well as a minimum mass below which stimulated capture only becomes significant after the onset of structure formation. The allowed range of axion masses shifts towards smaller values for heavier PBHs, keeping the dimensionless mass coupling $\alpha\sim 10^{-9}-10^{-5}$. There is also a minimum self-coupling value for efficient cloud growth, or equivalently a maximum value of $f_a$. This shifts towards larger $f_a$ for heavier PBHs, i.e.~it is easier to grow axion clouds around heavier PBHs.  

In Figure~\ref{fig:alps_region} we also plot current observational constraints on ALPs, showing that there is a viable PBH and axion mass range for cloud growth through stimulated capture. Note that stripping processes only constrain the parameter space for the largest values of the PBH mass considered, and only for small values of $f_a$ that are already excluded by current observational constraints.

It is interesting to point out that constraints from black hole superradiance (the gray region in Figure~\ref{fig:alps_region}), which are obtained through spin measurements of known stellar mass black holes, correspond to a parametric region where stimulated capture is inefficient. Both superradiance and stimulated capture correspond to viable ways of forming bound axion clouds around black holes, with exponential growth rates. Superradiance, on the one hand, is powered by the black hole's rotational energy, with self-interactions quenching cloud growth through collisional ionization \cite{Baryakhtar:2020gao}; stimulated capture, on the other hand, requires sufficiently large self-scattering rates to become significant. The two mechanisms also operate in different ranges of the dimensionless mass coupling $\alpha$: whereas large $\alpha$ promotes growth for spinning BHs, it promotes absorption in the Schwarzschild case. This explains why, for instance, $10^{-12}$ eV-axion cloud formation may occur through superradiance for spinning stellar mass BHs (at large $f_a)$ and through stimulated capture for non-spinning planetary-mass PBHs (at small $f_a)$. 

While in Figure~\ref{fig:alps_region} we show the line corresponding to the QCD axion, suggesting a viable region for cloud formation around the smallest PBHs considered\footnote{Note that PBHs lighter than $\sim 10^{12}$ kg are expected to evaporate through Hawking emission before the present day, so we focus on heavier PBHs.}, this case requires a dedicated analysis, as we discuss in the next sub-section.

%=========================================================
\subsection{QCD axion}\label{subsec:QCDaxion}
% =========================================================

For the QCD axion, the zero-temperature mass is not an independent parameter but is fixed by the decay constant according to:
\begin{equation}
\mu_0 \;\simeq\; 6\times 10^{-3}\,{\rm eV}\,\left(\frac{10^{9}\,{\rm GeV}}{f_a}\right)\, .
\label{eq:qcd_mass_fa_relation}
\end{equation}
A further important difference with respect to the constant-mass ALP case is that prior to the QCD crossover ($T\sim \Lambda_{\rm{QCD}}\simeq 150$ MeV) the axion mass evolves with temperature. This modifies the onset of oscillations, the time dependence of $\alpha(T)$, the post-oscillation velocity redshifting, and the
energy-density evolution. We adopt a power-law parametrization,
\begin{equation}
\mu(T)=\mu_0\,
\begin{cases}
\left(\dfrac{\Lambda_{\rm QCD}}{T}\right)^{n/2}, & T>\Lambda_{\rm QCD},\\[6pt]
1, & T\le \Lambda_{\rm QCD},
\end{cases}
\label{eq:mass_qcd_piecewise_compact}
\end{equation}
where we will assume $n=8$ throughout \cite{Marsh:2017hbv}. Since $\alpha\propto \mu$, we have,
\begin{equation}
\alpha(T)=\alpha_0\,
\begin{cases}
\left(\dfrac{\Lambda_{\rm QCD}}{T}\right)^{4}, & T>\Lambda_{\rm QCD},\\[6pt]
1, & T\le \Lambda_{\rm QCD},
\end{cases}
\qquad \alpha_0\equiv \frac{\mu_0 M_{\rm PBH}}{\mpl^{2}}\,.
\label{eq:alpha_qcd_piecewise_compact}
\end{equation}
Coherent oscillations start once the Hubble rate becomes comparable to the temperature-dependent axion mass, $H(T_{\rm osc})\simeq \mu(T_{\rm osc})$. During radiation domination and for $T_{\rm osc}>\Lambda_{\rm QCD}$ one finds,
\begin{equation}
T_{\rm osc}\simeq 3.6\,{\rm GeV}\,
\left(\frac{\mu_0}{10^{-3}\,{\rm eV}}\right)^{1/6}\,
g_*^{-1/12}(T_{\rm osc})\,.
\label{eq:TOSCqcd}
\end{equation}
Unlike the constant-mass ALP case, where the threshold \(t_*\) can in principle occur during radiation or matter domination, all viable QCD-axion cloud formation scenarios reach the stimulated capture regime in the radiation epoch, with typically $T_*$ near or above the QCD scale.
It is therefore sufficient to restrict the QCD discussion to radiation domination. The physically relevant distinction then arises instead from the behavior of \(\mu(T)\) across the QCD crossover, which leads to two qualitatively different regimes: \(T_*>\Lambda_{\rm QCD}\) and \(T_*<\Lambda_{\rm QCD}\):

% ------------------------------------------------------------------------------------------------------------------

\begin{enumerate}

\item[\bf 1.] {\bf Axion cloud formation before the QCD crossover} ($T_*> \Lambda_{\rm QCD}$)

In this regime,
\begin{equation}
\mu(T_*)=\mu_0\left(\frac{\Lambda_{\rm QCD}}{T_*}\right)^{4},
\qquad
\alpha(T_*)=\alpha_0\left(\frac{\Lambda_{\rm QCD}}{T_*}\right)^{4}\, .
\end{equation}
The physical momentum redshifts as $p\propto a^{-1}$, but $v=p/\mu(T)$, so the varying mass modifies the standard scaling $v\propto a^{-1}$. After oscillations begin one may write,
\begin{equation}
v(T)=v_{\rm osc}\,\frac{a_{\rm osc}}{a(T)}\,\frac{\mu(T_{\rm osc})}{\mu(T)}\simeq \left(\frac{T}{T_{\rm osc}}\right)^{5}
\left(\frac{g_{*s}(T)}{g_{*s}(T_{\rm osc})}\right)^{1/3}\, ,
\label{eq:v_qcd_master}
\end{equation}
where we assumed $v_{\rm osc}\sim1$. Imposing the velocity threshold condition leads to,
\begin{equation}
T_*
\simeq
0.45\,{\rm GeV}\,
\left(\frac{\alpha_0}{10^{-6}}\right)^{1/9}
\left(\frac{\mu_0}{1\ {\rm eV}}\right)^{5/54}\,
\frac{g_{*s}^{1/27}(T_{\rm osc})}{g_{*s}^{1/27}(T_*)\,g_*^{5/108}(T_{\rm osc})}\, .
\label{eq:Tstar_caseI}
\end{equation}

A second important difference relative to constant-mass ALPs is the evolution of the energy density. The comoving number density is conserved, $n_a a^3=\mathrm{const.}$, but $\rhoa(T)=\mu(T)\,n_a$; hence, for $T>\Lambda_{\rm QCD}$,
\begin{equation}
\rhoa(T)=\rhoa^0\,a^{-3}\,\frac{\mu(T)}{\mu_0}\,.
\label{eq:rhoa_caseI_scaling}
\end{equation}
Combining these ingredients with the stimulated capture rate and evaluating the growth criterion at the threshold yields,
\begin{equation}
\left.\frac{\Gamma_{1}}{H}\right|_{T_*}
\simeq 9\times 10^5\,f_{\rm DM}^{2}\,
\left(\frac{10^{-9}}{\alpha_0}\right)^{2/9}
\left(\frac{\mu_0}{10^{-3}\,{\rm eV}}\right)^{67/27}\,
\frac{g_{*s}^{16/27}(T_{\rm osc})\, g_{*s}^{20/27}(T_*)}{g_*^{20/27}(T_{\rm osc})}\, .
\label{eq:GammaH_caseI_muonly}
\end{equation}

% ---------------------------------------------------------

\item[\bf 2.] {\bf Axion cloud formation after the QCD crossover} ($T_*< \Lambda_{\rm QCD}$)

%---------------------------------------------------------

If the stimulated capture threshold is reached after the axion mass has saturated, then $\mu(T_*)=\mu_0$ and $\alpha(T_*)=\alpha_0$.
After the QCD crossover the axion condensate velocity redshifts as for non-relativistic matter with constant mass, $v\propto a^{-1}$, and can be written as,
\begin{equation}
v(T)
=v_{\rm QCD}\,\frac{T}{\Lambda_{\rm QCD}}
\left(\frac{g_{*s}(T)}{g_{*s}(\Lambda_{\rm QCD})}\right)^{1/3}\,,
\,
v_{\rm QCD}\simeq \left(\frac{\Lambda_{\rm QCD}}{T_{\rm osc}}\right)^5
\left(\frac{g_{*s}(\Lambda_{\rm QCD})}{g_{*s}(T_{\rm osc})}\right)^{1/3}\,.
\label{eq:v_qcd_caseII_general}
\end{equation}
Imposing the velocity threshold condition yields,
\begin{equation}
T_*
\simeq 7.4\ {\rm MeV}\,
\left(\frac{\mu_0}{10^{-3}\,{\rm eV}}\right)^{5/6}
\left(\frac{\alpha_0}{10^{-9}}\right)\,
\left(\frac{g_{*s}(T_{\rm osc})}{g_{*s}(T_*)}\right)^{1/3}
g_*^{-5/12}(T_{\rm osc})\,.
\label{eq:Tstar_caseII}
\end{equation}
Since the axion mass is already constant at the threshold for stimulated capture, the axion energy density redshifts as $\rhoa\propto a^{-3}$, as in the ALP case.
Evaluating the growth criterion at $T_*$ then gives,
\begin{equation}
\left.\frac{\Gamma_{1}}{H}\right|_{T_*}
\simeq
10^3f_{\rm DM}^{2}\,
\left(\frac{\alpha_0}{10^{-9}}\right)^{2}
\left(\frac{\mu_0}{10^{-3}\,{\rm eV}}\right)^{13/3}\,
\frac{g_{*s}^{4/3}(T_{\rm osc})}{g_*^{5/3}(T_{\rm osc})}\,.
\label{eq:GammaH_caseII_muonly}
\end{equation}
\end{enumerate}

\begin{figure}[h]
    \centering
\includegraphics[width=0.95\textwidth]{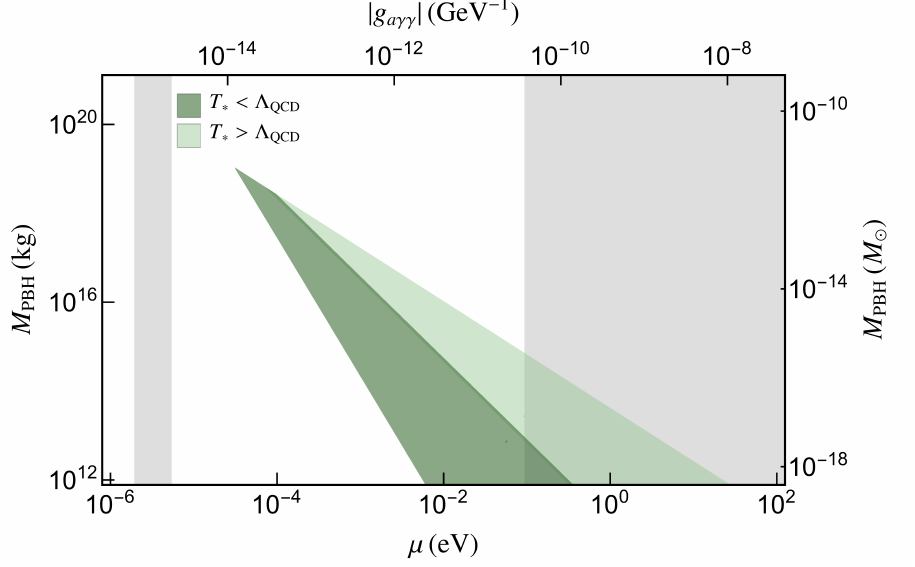}
    \caption{\small
Parameter space for PBH-seeded stimulated growth of QCD-axion clouds in the $(\mu,M_{\rm PBH})$ plane.
Colored regions satisfy the same physical requirements as in Fig.~\ref{fig:alps_region} for stimulated growth and stability. Because the QCD axion mass is temperature dependent when $T_*>\Lambda_{\rm QCD}$, the allowed region splits into two cases:
a varying-mass regime where the stimulated threshold occurs at $T_*>\Lambda_{\rm QCD}$ (light green), and a constant-mass regime where
$T_*<\Lambda_{\rm QCD}$ (dark green).
Gray vertical bands indicate regions excluded by current bounds on $|g_{a\gamma\gamma}|$ mapped onto the QCD-axion trajectory \cite{AxionLimits,Gorghetto:2022ikz}.}
\label{fig:qcd_region}
\end{figure}

In addition to these modified growth criteria, we also impose, as for the ALP case, the cloud stability conditions from PBH reabsorption and environmental stripping, which are the same as for ALPs. The viable parametric regions are shown in Figure ~\ref{fig:qcd_region}, with the two cases shown in different colors for clarity. 

The obtained axion and PBH mass range agrees roughly with what we obtained for constant mass ALPs (the QCD axion line in Figure \ref{fig:alps_region}). For too large axion/PBH masses the clouds form but are reabsorbed by the PBH before the present day, while for small axion masses (large $f_a$) self-interactions are too weak for stimulated capture to be efficient. This nevertheless leaves a broad viable window around $\mu\sim 10^{-5}$--$10^{-1}\,\mathrm{eV}$ and $M_{\rm PBH}\sim 10^{12}$--$10^{19}\,\mathrm{kg}$, consistent with observational constraints. 

Taken together, Figures~\ref{fig:alps_region} and \ref{fig:qcd_region} show that stimulated capture can populate macroscopic axion(-like) clouds around PBHs in regions of parameter space that are not excluded by existing axion searches. The surviving windows therefore provide well-motivated targets for the observational strategies discussed in Section~\ref{sec:Detection}.

% =========================================================
\section{Detectability of PBH-axion clouds}\label{sec:Detection}
% =========================================================

Let us now discuss the detectability of spherical axion/ALP clouds around PBHs in the mass ranges obtained in the previous section. We will first focus on axion-to-photon conversion in strong magnetic fields, which can lead to narrow-band electromagnetic emission. Second, we will discuss the potential effects of extended axion clouds in PBH microlensing events, discussing the parametric regimes where the corresponding light curves may depart from the standard point-lens Paczy\'nski form. These two types of signatures correspond, as we will see, to the high (light) and low (heavy) axion (PBH) mass regimes, respectively.

\subsection{Axion-to-photon conversion in neutron star magnetic fields}
\label{subsec:magnetar_conversion}

Axions and ALPs generically couple to electromagnetism through,
\begin{equation}
\mathcal{L}
\supset
-\frac{1}{4}\,g_{a\gamma\gamma}\,a\,F_{\mu\nu}\tilde F^{\mu\nu},
\qquad
g_{a\gamma\gamma}
\equiv
\frac{\alpha_{\rm EM}}{2\pi f_a}\,c_\gamma \, ,
\label{eq:axion_photon_coupling}
\end{equation}
where \(\alpha_{\rm EM}\) is the electromagnetic fine-structure constant, \(c_\gamma\) is an $\mathcal{O}(1)$
model-dependent dimensionless coefficient, \(F_{\mu\nu}\) is the electromagnetic field-strength
tensor, and \(\tilde F^{\mu\nu}\equiv \epsilon^{\mu\nu\rho\sigma}F_{\rho\sigma}/2\) is its dual. A potentially efficient electromagnetic signature can arise when a PBH-seeded axion cloud
crosses a strongly magnetized plasma environment, such as the magnetosphere of a neutron star.
In this case the cloud is not treated as an intrinsically luminous source. Rather, it acts as
a reservoir of non-relativistic axions which may resonantly convert into photons in the external
magnetic field. The resonance condition is set by the equality between the local plasma frequency and the
axion mass,
\begin{equation}
\omega_p(r_c)
\equiv
\left(
\frac{4\pi\alpha_{\rm EM} n_e(r_c)}{m_e}
\right)^{1/2}
=
\mu \, ,
\label{eq:resonance_condition_magnetar}
\end{equation}
where \(r_c\) is the conversion radius and \(n_e(r)\) is the local electron number density.
We adopt a benchmark neutron star with radius \(R_{\rm NS}=12\,{\rm km}\), mass
\(M_{\rm NS}=1.4\,M_\odot\), spin period \(\tau=0.1\,{\rm s}\) and surface magnetic field
\(B_0=10^{9}\,{\rm T}\). We will consider, for simplicity, an aligned dipolar magnetic field \cite{Song:2024rru},
\begin{equation}
B(r)
=
B_0
\left(
\frac{R_{\rm NS}}{r}
\right)^3 \, ,
\label{eq:dipole_B_magnetar}
\end{equation}
noting that misalignment effects will not change the order of magnitude of our results.
For the plasma density we take a Goldreich--Julian profile multiplied by an effective pair
multiplicity factor \(\kappa\) \cite{Witte:2021arp},
\begin{equation}
n_e(r)
=
\kappa\,
\frac{4\pi\epsilon_0 B_0}{e\,\tau}
\left(
\frac{R_{\rm NS}}{r}
\right)^3\simeq
7\times 10^{14}\,{\rm cm}^{-3}
\left(
\frac{\kappa}{10^2}
\right)\left(
\frac{B_0}{10^{9}\,{\rm T}}
\right)\left(\frac{0.1 \, {\rm s}}{\tau}\right)
\left(
\frac{R_{\rm NS}}{r}
\right)^3 \, ,
\label{eq:GJ_density_general}
\end{equation}
which yields the plasma frequency,
\begin{equation}
\omega_p(r)
\simeq
10^{-3}\,{\rm eV}
\left(
\frac{\kappa}{10^2}
\right)^{1/2}\left(
\frac{B_0}{10^{9}\,{\rm T}}
\right)^{1/2}\left(
\frac{0.1\,{\rm s}}{\tau}
\right)^{1/2}
\left(
\frac{R_{\rm NS}}{r}
\right)^{3/2}\, .
\label{eq:wp_numerical}
\end{equation}
The conversion radius at which resonance occurs is, thus,
\begin{equation}
r_c
\simeq
12\,{\rm km}
\left(
\frac{\kappa}{10^2}
\right)^{1/3}\left(
\frac{B_0}{10^{9}\,{\rm T}}
\right)^{1/3}\left(
\frac{0.1\,{\rm s}}{\tau}
\right)^{1/3}
\left(
\frac{10^{-3}\,{\rm eV}}{\mu}
\right)^{2/3}\, .
\label{eq:rc_general}
\end{equation}
The resonance condition therefore defines an approximately spherical conversion shell at radius $r=r_c$, surrounding the neutron star. Throughout the following discussion, we refer to this physical surface as the resonant shell.
The Goldreich--Julian density represents the minimum charge density required to sustain
corotation, while electron--positron pair cascades may increase the total plasma density by
several orders of magnitude. Recent axion-conversion studies have considered plasma
densities $n_e/n_{\rm GJ}=1,10^2$, and $10^4$~\cite{Song:2024rru}. We therefore adopt
$\kappa=10^2$ as a representative benchmark at the lower end of the pair-enhanced range.
In this analysis we focus only on points satisfying \(r_c>R_{\rm NS}\) corresponding to conversion outside the neutron star surface. Near resonance, the axion-to-photon conversion probability may be estimated using the Landau--Zener expression \cite{Carenza:2023nck},
\begin{align}
P_{a\to\gamma}^{\rm res}(r_c)
&=
1-
\exp\left[
-
\frac{\pi g_{a\gamma\gamma}^2 B_T^2(r_c)}
{v_\perp\,\mu\,
\left|
\partial_\ell \ln\omega_p^2
\right|_{r_c}}
\right]
\nonumber\\[0.5em]
&\simeq
\frac{3\times 10^{-4}}{v_\perp}
\left(
\frac{10^8\,{\rm GeV}}{f_a}
\right)^2
\left(
\frac{10^2}{\kappa}
\right)^{5/3}
\left(
\frac{B_0}{10^9\,{\rm T}}
\right)^{1/3}
\left(
\frac{\tau}{0.1\,{\rm s}}
\right)^{5/3}
\left(
\frac{\mu}{10^{-3}\,{\rm eV}}
\right)^{7/3}
c_\gamma^2 \, .
\label{eq:Pres_LZ}
\end{align}
Here \(B_T\) is the magnetic-field component transverse to the axion trajectory, and
\(v_\perp\) is the local velocity of the axion flow relative to the magnetospheric plasma,
projected along the normal to the resonant surface. For a central, approximately radial crossing one may take\footnote{Note that for non-radial crossings the signal may be suppressed by geometrical factors, although with the same order of magnitude.},
\begin{equation}
v_\perp \simeq v_c(r_c)
=
\left[
v_{\rm DM}^2+
v_{\rm esc}^2(r_c)
\right]^{1/2},
\qquad
v_{\rm esc}(r_c)=\left(\frac{2GM_{\rm NS}}{r_c}\right)^{1/2} \, ,
\end{equation}
up to geometric factors of order unity. In the second line of Eq. (\ref{eq:Pres_LZ}) we used the power-law plasma profile above, for which,
\begin{equation}
\left|
\partial_\ell\ln\omega_p^2
\right|_{r_c}
\simeq
\frac{3}{r_c} \, ,
\end{equation}
for approximately radial trajectories, and we took the small-probability limit (valid when the exponent is much smaller than unity). 

The conversion probability derived above gives the probability for a single axion to convert during one crossing of the resonant shell.
To assess the observational relevance of the
signal, one must also estimate how often PBH-seeded clouds encounter neutron star
magnetospheres. The local PBH flux at the galactic Solar radius is,
\begin{equation}
\Phi_{\rm PBH}^{\rm loc}
=
\frac{f_{\rm PBH}\rho_\odot}{M_{\rm PBH}}v_0
\simeq
2\times 10^{-32}\,{\rm m}^{-2}{\rm s}^{-1}
\left(
\frac{10^{14}\,{\rm kg}}{M_{\rm PBH}}
\right)
\left(\frac{f_{\rm PBH}}{10^{-2}}\right)
\, ,
\label{eq:PBH_flux_local}
\end{equation}
where $\rho_\odot\simeq0.4\,{\rm GeV\,cm^{-3}}$ and $v_0\simeq220\,{\rm km\,s^{-1}}$ denote the DM density and the
characteristic PBH--neutron star relative velocity at the Solar radius
\cite{deSalas:2020hbh,McMillan:2016}. The corresponding local-density crossing rate for one neutron star is,
\begin{equation}
\Gamma_{\rm PBH,loc}^{\rm NS}
=
\Phi_{\rm PBH}^{\rm loc}\,
\pi r_c^2
\left[
1+
\frac{v_{\rm esc}^2(r_c)}{v_0^2}
\right]\, .
\end{equation}
The factor in brackets accounts for gravitational focusing: a PBH with
asymptotic impact parameter larger than $r_c$ can still be focused onto
an orbit whose pericentre lies inside the resonant shell. Since
$v_{\rm esc}(r_c)\gg v_0$ for the benchmark parameters, the local rate is
in the strong-focusing regime and scales approximately as
$\Gamma_{\rm PBH,loc}^{\rm NS}\propto r_c/v_0$. Numerically,
\begin{align}
\Gamma_{\rm PBH,loc}^{\rm NS}
\simeq
10^{-10}\,{\rm yr}^{-1}
\left(
\frac{\kappa}{10^2}
\right)^{1/3}
\left(
\frac{B_0}{10^9\,{\rm T}}
\right)^{1/3}
\left(
\frac{0.1\,{\rm s}}{\tau}
\right)^{1/3}
\left(
\frac{10^{-3}\,{\rm eV}}{\mu}
\right)^{2/3}
\left(
\frac{10^{14}\,{\rm kg}}{M_{\rm PBH}}
\right)
\left(
\frac{f_{\rm PBH}}{10^{-2}}
\right)\, .
\label{eq:single_magnetar_rate_mu}
\end{align}
A more realistic Galactic rate is obtained by averaging the PBH flux over
the spatial distribution of neutron stars in the Milky Way. Let
$n_{\rm NS}^{\rm Gal}(\mathbf x)$ be the Galactic neutron star number
density, normalized as,
\begin{equation}
\int d^3x\,n_{\rm NS}^{\rm Gal}(\mathbf x)
=
\mathcal N_{\rm NS}\, .
\end{equation}
At position $\mathbf x$, we denote by $u(\mathbf x)$ the PBH--neutron star
relative velocity far from the neutron star, but at that Galactocentric
position. The Galactic crossing rate
is then,
\begin{equation}
\Gamma_{\rm PBH}^{\rm Gal}
\simeq
\frac{f_{\rm PBH}}{M_{\rm PBH}}
\int d^3x\,
n_{\rm NS}^{\rm Gal}(\mathbf x)
\rho_{\rm DM}^{\rm Gal}(\mathbf x)
\pi r_c^2\frac{v_{\rm esc}^2(r_c)}{u(\mathbf x)}\, ,
\label{eq:galactic_profile_averaged_rate}
\end{equation}
where we considered the strong-focusing limit, $v_{\rm esc}(r_c)\gg u(\mathbf x)$.
Equivalently, we write,
\begin{equation}
\Gamma_{\rm PBH}^{\rm Gal}
=
\mathcal B_{\rm Gal}\,
\mathcal N_{\rm NS}\,
\Gamma_{\rm PBH,loc}^{\rm NS}\, ,
\label{eq:galactic_rate_boost_def}
\end{equation}
where,
\begin{equation}
\mathcal B_{\rm Gal}
=
\frac{1}{\mathcal N_{\rm NS}}
\int d^3x\,
n_{\rm NS}^{\rm Gal}(\mathbf x)
\frac{\rho_{\rm DM}^{\rm Gal}(\mathbf x)}{\rho_\odot}
\frac{v_0}{u(\mathbf x)} \, .
\label{eq:galactic_boost_general}
\end{equation}
For illustration, we estimate $\mathcal B_{\rm Gal}$ using a simple Milky-Way toy model. We take an NFW DM profile
\cite{Navarro:1996gj,Navarro:1997iw},
\begin{equation}
\rho_{\rm DM}^{\rm Gal}(r)
=
\frac{\rho_s}{(r/r_s)(1+r/r_s)^2},
\qquad
\rho_s
=
\rho_\odot
\frac{R_0}{r_s}
\left(
1+\frac{R_0}{r_s}
\right)^2 \, ,
\label{eq:NFW_profile_galactic}
\end{equation}
where $R_0=8.2\,{\rm kpc}$ is the Solar Galactocentric radius and $r_s=20\,{\rm kpc}$ represents the scale radius of the NFW halo obtained in dynamical mass models of the Milky Way \cite{McMillan:2016}. For the neutron star
population we use an effective spherical Dehnen profile
\cite{Hernquist:1990be,Dehnen:1993},
\begin{equation}
n_{\rm NS}^{\rm Gal}(r)
=
\mathcal N_{\rm NS}
\frac{3-\gamma_{\rm NS}}{4\pi}
\frac{a_{\rm NS}}
{r^{\gamma_{\rm NS}}(r+a_{\rm NS})^{4-\gamma_{\rm NS}}}\, ,
\label{eq:NS_Dehnen_profile_galactic}
\end{equation}
which is normalized to $\mathcal N_{\rm NS}$. Here $a_{\rm NS}=3\,{\rm kpc}$ is a scale radius of the assumed neutron star spatial distribution which we take as a representative kiloparsec-scale transition radius between the inner cusp and the outer fall off \cite{McMillan:2016}. Furthermore,
$\gamma_{\rm NS}=1$ corresponds to the Hernquist member of the Dehnen family \cite{Hernquist:1990be,Dehnen:1993}. This is not
intended as a detailed reconstruction of the Galactic neutron star
population, but as a controlled effective model for the enhancement due to the concentration of compact stellar remnants towards the inner
Galaxy. For the velocity profile, we use a phenomenological prescription based
on the circular-speed scaling of an NFW halo \cite{Navarro:1996gj}. For a spherical NFW profile, the
enclosed mass is proportional to,
\begin{equation}
\mathcal M(r/r_s)
=
\ln(1+r/r_s)-\frac{r/r_s}{1+r/r_s}\, .
\end{equation}
Thus the halo circular speed scales as
$v_c^2(r)\propto \mathcal M(r/r_s)/(r/r_s)$. We use this scaling only as a proxy for the PBH--neutron star relative velocity in the host galaxy,
and normalize it to the local reference value $v_0$,
\begin{equation}
u(r)
=
\left[
u_{\rm min}^2
+
\left(
v_0^2-u_{\rm min}^2
\right)
\frac{\mathcal M(r/r_s)/(r/r_s)}
{\mathcal M(R_0/r_s)/(R_0/r_s)}
\right]^{1/2}\, .
\label{eq:velocity_profile_galactic}
\end{equation}
The velocity floor $u_{\rm min}$ regularizes the inner-halo behaviour of
the NFW-only circular-speed scaling and parametrizes the fact that the
central relative velocity is affected by the stellar bulge, the disk, the
central black hole, and neutron star natal kicks
\cite{McMillan:2016,Eilers:2018,Hobbs:2005yx,Sartore:2010}. In the
numerical estimate we use
$u_{\rm min}=50\text{--}100\,{\rm km\,s^{-1}}$ giving $\mathcal B_{\rm Gal}\sim 10$. For an effective Galactic population of $\mathcal N_{\rm NS}=10^9$ neutron
stars \cite{Treves:1999,Popov:2000,Sartore:2010}, the Galactic crossing
rate is therefore,
\begin{align}
\Gamma_{\rm PBH}^{\rm Gal}
\simeq
&1\,{\rm yr}^{-1}
\left(
\frac{\mathcal B_{\rm Gal}}{10}
\right)
\left(
\frac{\mathcal N_{\rm NS}}{10^9}
\right)
\left(
\frac{\kappa}{10^2}
\right)^{1/3}
\left(
\frac{B_0}{10^9\,{\rm T}}
\right)^{1/3}
\left(
\frac{0.1\,{\rm s}}{\tau}
\right)^{1/3}\times \nonumber \\
& \times
\left(
\frac{10^{-3}\,{\rm eV}}{\mu}
\right)^{2/3}
\left(
\frac{10^{14}\,{\rm kg}}{M_{\rm PBH}}
\right)
\left(
\frac{f_{\rm PBH}}{10^{-2}}
\right)\, .
\end{align}
Requiring at least one event every ten years,
$\Gamma_{\rm PBH}^{\rm Gal}\gtrsim0.1\,{\rm yr}^{-1}$, gives,
\begin{equation}
B_0
\gtrsim
10^6\,{\rm T}
\left(
\frac{\tau}{0.1\,{\rm s}}
\right)
\left(
\frac{\mu}{10^{-3}\,{\rm eV}}
\right)^2
\left(
\frac{10}{\mathcal B_{\rm Gal}}
\right)^{3}
\left(
\frac{10^9}{\mathcal N_{\rm NS}}
\right)^3
\left(
\frac{M_{\rm PBH}}{10^{14}\,{\rm kg}}
\right)^3
\left(
\frac{10^{-2}}{f_{\rm PBH}}
\right)^3
\left(
\frac{10^2}{\kappa}
\right)
 \, .
\label{eq:B0_min_event_rate}
\end{equation}

Having fixed the encounter rate, we now estimate the brightness of an individual event.
The relevant quantity is the axion mass flux through the resonant shell, multiplied by the conversion probability. The resulting signal depends on the ratio between the cloud size $R_\star$ and the shell radius $r_c$. If $R_\star\ll r_c$, the curvature of the shell is negligible across the spatial extent of the cloud, and each entry or exit crossing may be treated using a local tangent-plane approximation. If instead $R_\star\gg r_c$, the cloud extends over the entire resonant shell, and only the fraction of the cloud intersecting the projected shell area contributes at a given time, leading to a geometrical suppression, as illustrated in Figure \ref{fig:cloud_geometry_regimes}. The PBH-seeded axion
cloud is described by the density profile,
\begin{equation}
\rho_{\rm cloud}(r)
=
\frac{M_{\rm cloud}}{\pi R_\star^3}
e^{-2r/R_\star}\, ,
\label{eq:cloud_profile}
\end{equation}
which corresponds to the $1s$ Hydrogenic ground state. For the cloud mass we adopt, as an optimistic benchmark, the maximal cloud mass reached before
self-interactions invalidate the perturbative regime or saturate the growth \cite{Budker:2023sex},
\begin{equation}
M_{\rm cloud}\simeq 2\times 10^{11}~{\rm kg}\;
\left(\frac{10^{-9}}{\alpha}\right)^2
\left(\frac{f_a}{10^{8}~{\rm GeV}}\right)^2
\left(\frac{M_{\rm PBH}}{10^{14}~{\rm kg}}\right) \, .
\label{eq:Mcrit}
\end{equation}
The ratio between cloud size and conversion radius is,
\begin{equation}
\frac{R_\star}{r_c}
\simeq
4\times10^1
\left(
\frac{10^2}{\kappa}
\right)^{1/3}
\left(
\frac{\tau}{0.1\,{\rm s}}
\right)^{1/3}
\left(
\frac{10^9\,{\rm T}}{B_0}
\right)^{1/3}
\left(
\frac{10^{-3}\,{\rm eV}}{\mu}
\right)^{4/3}
\left(
\frac{10^{14}\,{\rm kg}}{M_{\rm PBH}}
\right) \, .
\label{eq:Rstar_over_rc}
\end{equation}
Thus, for the benchmark parameters and \(\mu\simeq 10^{-3}\,{\rm eV}\), the cloud is large
relative to the resonant shell. At higher axion masses, \(R_\star/r_c\propto \mu^{-4/3}\)
decreases rapidly and the cloud approaches the compact regime. In this limit, let \(s\) denote the signed normal distance between the cloud
centre and the local tangent plane to the resonant shell. The point \(O\)
is the orthogonal projection of the cloud centre onto this plane, while
\(\mathbf{x}_\perp\) denotes the displacement within the tangent plane. The cloud column density crossing the shell is,
\begin{equation}
\Sigma_{\rm cl}(s)
=
\int d^2x_\perp\,
\rho_{\rm cl}
\left(
\sqrt{s^2+x_\perp^2}
\right)
=
\frac{M_{\rm cloud}}{2R_\star}
e^{-2|s|/R_\star}
\left(
1+\frac{2|s|}{R_\star}
\right)\, .
\label{eq:cloud_column_density}
\end{equation}
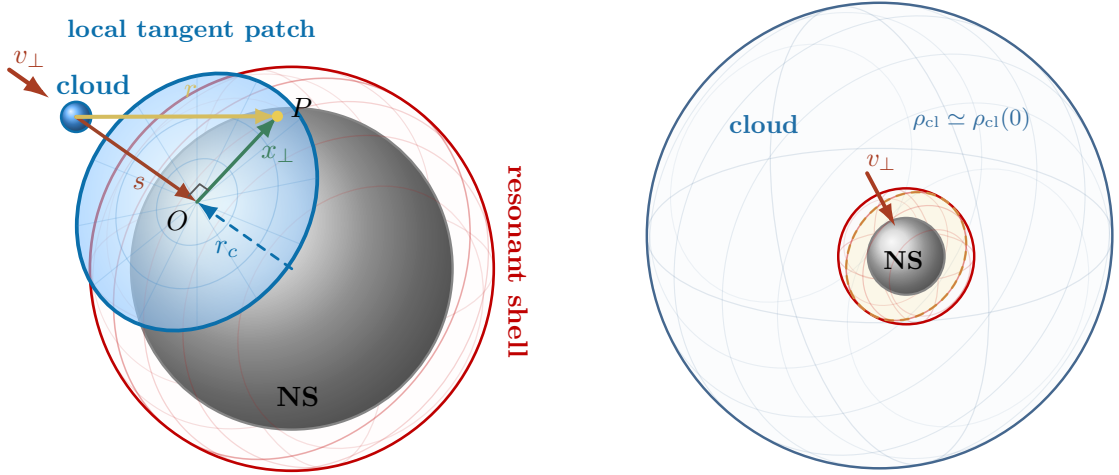
\begin{figure}[t]
\centering

%================================================
% PANEL (a): COMPACT CLOUD, R_* << r_c
%================================================
\begin{subfigure}[t]{0.48\linewidth}
\centering
\resizebox{\linewidth}{!}{%
\tdplotsetmaincoords{68}{62}

\begin{tikzpicture}[
    tdplot_main_coords,
    >=Latex,
    line cap=round,
    line join=round,
    every node/.style={font=\small}
]

%------------------------------------------------
% Geometrical scale
%------------------------------------------------
\def\rc{2.55}

%------------------------------------------------
% Fundamental points
%------------------------------------------------
\coordinate (C)  at (-3.25,0,0);
\coordinate (O)  at (0,0,0);
\coordinate (P)  at (0,1.15,0.95);
\coordinate (NS) at (2.55,0,0);

%================================================
% TRANSPARENT THREE-DIMENSIONAL RESONANT SHELL
%================================================

%------------------------------------------------
% Transparent projected spherical envelope
%------------------------------------------------
\node[
    circle,
    draw=red!75!black,
    fill=red!18,
    fill opacity=0.045,
    line width=0.95pt,
    minimum size=5.10cm,
    inner sep=0pt
] at (NS) {};

%------------------------------------------------
% Rear latitude sections
%------------------------------------------------
\foreach \dx/\op in {
    -2.00/0.09,
    -1.45/0.11,
    -0.80/0.13,
     0.00/0.15
}{
    \pgfmathsetmacro{\xx}{2.55+\dx}
    \pgfmathsetmacro{\rr}{sqrt(\rc*\rc-\dx*\dx)}

    \begin{scope}[canvas is yz plane at x=\xx]
        \draw[
            red!72!black,
            opacity=\op,
            line width=0.42pt
        ]
        (0,0) circle (\rr);
    \end{scope}
}

%------------------------------------------------
% Rear meridians
%------------------------------------------------
\begin{scope}[canvas is xz plane at y=0]
    \draw[
        red!72!black,
        opacity=0.12,
        line width=0.48pt
    ]
    (2.55,0) circle (\rc);
\end{scope}

\begin{scope}[canvas is xy plane at z=0]
    \draw[
        red!72!black,
        opacity=0.12,
        line width=0.48pt
    ]
    (2.55,0) circle (\rc);
\end{scope}

%------------------------------------------------
% Neutron star
%------------------------------------------------
\node[
    circle,
    shade,
    ball color=black!20,
    minimum size=4.06cm,
    inner sep=0pt,
    draw=black!48,
    line width=0.85pt
] at (NS) {};

%------------------------------------------------
% Front latitude sections
%------------------------------------------------
\foreach \dx/\op in {
    0.65/0.17,
    1.25/0.15,
    1.85/0.12
}{
    \pgfmathsetmacro{\xx}{2.55+\dx}
    \pgfmathsetmacro{\rr}{sqrt(\rc*\rc-\dx*\dx)}

    \begin{scope}[canvas is yz plane at x=\xx]
        \draw[
            red!78!black,
            opacity=\op,
            line width=0.48pt
        ]
        (0,0) circle (\rr);
    \end{scope}
}

%------------------------------------------------
% Principal shell great circle
%------------------------------------------------
\begin{scope}[canvas is yz plane at x=2.55]
    \draw[
        red!78!black,
        opacity=0.23,
        line width=0.62pt
    ]
    (0,0) circle (\rc);
\end{scope}

%================================================
% LOCAL TANGENT PATCH
%================================================
\begin{scope}[canvas is yz plane at x=0]

    \shade[
        inner color=Cerulean!8,
        outer color=RoyalBlue!45,
        opacity=0.60
    ]
    (0,0) circle (1.72);

    \draw[
        RoyalBlue!85!black,
        line width=1.35pt
    ]
    (0,0) circle (1.72);

    \foreach \rho in {0.58,1.16}{
        \draw[
            RoyalBlue!70,
            opacity=0.28,
            line width=0.45pt
        ]
        (0,0) circle (\rho);
    }

    \foreach \ang in {0,30,...,150}{
        \draw[
            RoyalBlue!70,
            opacity=0.20,
            line width=0.4pt
        ]
        (\ang:1.72) -- (\ang+180:1.72);
    }

\end{scope}

%------------------------------------------------
% Axion cloud
%------------------------------------------------
\node[
    circle,
    shade,
    ball color=RoyalBlue!72,
    minimum size=0.38cm,
    inner sep=0pt,
    draw=RoyalBlue!85!black,
    line width=0.9pt
] at (C) {};

%------------------------------------------------
% Incoming velocity
%------------------------------------------------
\draw[
    -{Latex[length=3.4mm,width=2.2mm]},
    BrickRed!90!black,
    line width=1.55pt
]
(-5.05,0,0) -- (-3.95,0,0)
node[
    midway,
    above=4pt,
    text=BrickRed!90!black,
    inner sep=1.5pt
]
{$v_\perp$};

%================================================
% GEOMETRY
%================================================

% Shell radius
\draw[
    -{Latex[length=2.6mm]},
    RoyalBlue!85!black,
    dashed,
    line width=1.0pt
]
(NS) -- (O)
node[
    pos=0.50,
    below left=1pt,
    inner sep=0.8pt,
    text=RoyalBlue!85!black
]
{$r_c$};

% Normal separation
\draw[
    -{Latex[length=2.8mm]},
    BrickRed!82!black,
    line width=1.25pt
]
(C) -- (O)
node[
    pos=0.52,
    below=4pt,
    inner sep=1.2pt
]
{$s$};

% Transverse displacement
\draw[
    -{Latex[length=2.8mm]},
    ForestGreen!65!black,
    line width=1.25pt
]
(O) -- (P)
node[
    pos=0.56,
    right=5pt,
    inner sep=1.2pt
]
{$x_\perp$};

% Three-dimensional cloud radius
\draw[
    -{Latex[length=2.8mm]},
    Goldenrod!80!black,
    line width=1.45pt
]
(C) -- (P)
node[
    pos=0.57,
    above=5pt,
    sloped,
    inner sep=1.2pt
]
{$r$};

%------------------------------------------------
% Right-angle marker
%------------------------------------------------
\coordinate (A) at (-0.23,0,0);
\coordinate (B) at (-0.23,0.16,0.132);
\coordinate (D) at (0,0.16,0.132);

\draw[
    black!65,
    line width=0.65pt
]
(A) -- (B) -- (D);

%------------------------------------------------
% Points O and P
%------------------------------------------------
\fill[Goldenrod!90!black] (P) circle (2.0pt);

\node[
    anchor=west,
    inner sep=1.2pt
]
at ($(P)+(0,0.13,0.10)$)
{$P$};

\node[
    anchor=north east,
    inner sep=1pt
]
at ($(O)+(-0.03,-0.06,-0.08)$)
{$O$};

%------------------------------------------------
% Labels
%------------------------------------------------
\node[
    RoyalBlue!85!black,
    font=\small\bfseries,
    inner sep=1.5pt
]
at (-3.20,0.22,0.42)
{cloud};

\node[
    RoyalBlue!85!black,
    font=\footnotesize\bfseries
]
at (-0.15,0,2.28)
{local tangent patch};

\node[
    red!75!black,
    font=\small\bfseries,
    rotate=-90
]
at (2.90,3.05,-0.35)
{resonant shell};

\node[
    black,
    font=\small\bfseries,
    inner sep=1.5pt
]
at (2.90,-0.10,-1.60)
{NS};

\end{tikzpicture}%
}

\caption{
\(R_\star\ll r_c\): compact-cloud regime.
The resonant shell is locally approximated by a tangent plane.
}
\end{subfigure}%
\hfill
%================================================
% PANEL (b): EXTENDED CLOUD, R_* >> r_c
%================================================
\begin{subfigure}[t]{0.48\linewidth}
\centering
\resizebox{\linewidth}{!}{%
\tdplotsetmaincoords{68}{62}

\begin{tikzpicture}[
    tdplot_main_coords,
    >=Latex,
    line cap=round,
    line join=round,
    every node/.style={font=\small}
]

%------------------------------------------------
% Geometrical scales
%------------------------------------------------
\def\rcloud{3.15}
\def\rshell{0.92}
\def\rNS{0.52}

% Neutron-star position
\def\xNS{1.05}
\def\yNS{-0.15}
\def\zNS{0.10}

%------------------------------------------------
% Fundamental coordinates
%------------------------------------------------
\coordinate (CC) at (0,0,0);
\coordinate (NS) at (\xNS,\yNS,\zNS);

%================================================
% EXTENDED THREE-DIMENSIONAL CLOUD
%================================================

%------------------------------------------------
% Transparent spherical envelope
%------------------------------------------------
\node[
    circle,
    draw=RoyalBlue!58!black,
    fill=RoyalBlue!15,
    fill opacity=0.075,
    line width=0.95pt,
    minimum size=6.30cm,
    inner sep=0pt
] at (CC) {};

%------------------------------------------------
% Cloud cross sections
%------------------------------------------------
\foreach \dx/\op in {
    -2.55/0.08,
    -1.85/0.10,
    -1.05/0.12,
     0.00/0.14,
     1.05/0.12,
     1.85/0.10,
     2.55/0.08
}{
    \pgfmathsetmacro{\rr}{
        sqrt(\rcloud*\rcloud-\dx*\dx)
    }

    \begin{scope}[canvas is yz plane at x=\dx]
        \draw[
            RoyalBlue!68!black,
            opacity=\op,
            line width=0.43pt
        ]
        (0,0) circle (\rr);
    \end{scope}
}

%------------------------------------------------
% Two orthogonal cloud meridians
%------------------------------------------------
\begin{scope}[canvas is xz plane at y=0]
    \draw[
        RoyalBlue!65!black,
        opacity=0.15,
        line width=0.48pt
    ]
    (0,0) circle (\rcloud);
\end{scope}

\begin{scope}[canvas is xy plane at z=0]
    \draw[
        RoyalBlue!65!black,
        opacity=0.15,
        line width=0.48pt
    ]
    (0,0) circle (\rcloud);
\end{scope}

%================================================
% THREE-DIMENSIONAL RESONANT SHELL
%================================================

%------------------------------------------------
% Transparent shell envelope
%------------------------------------------------
\node[
    circle,
    draw=red!75!black,
    fill=red!18,
    fill opacity=0.045,
    line width=0.95pt,
    minimum size=1.84cm,
    inner sep=0pt
] at (NS) {};

%------------------------------------------------
% Rear shell cross sections
%------------------------------------------------
\foreach \dx/\op in {
    -0.72/0.12,
    -0.38/0.15,
     0.00/0.18
}{
    \pgfmathsetmacro{\xx}{\xNS+\dx}
    \pgfmathsetmacro{\rr}{
        sqrt(\rshell*\rshell-\dx*\dx)
    }

    \begin{scope}[canvas is yz plane at x=\xx]
        \draw[
            red!72!black,
            opacity=\op,
            line width=0.42pt
        ]
        (\yNS,\zNS) circle (\rr);
    \end{scope}
}

%------------------------------------------------
% Shell meridians
%------------------------------------------------
\begin{scope}[canvas is xz plane at y=\yNS]
    \draw[
        red!72!black,
        opacity=0.18,
        line width=0.48pt
    ]
    (\xNS,\zNS) circle (\rshell);
\end{scope}

\begin{scope}[canvas is xy plane at z=\zNS]
    \draw[
        red!72!black,
        opacity=0.18,
        line width=0.48pt
    ]
    (\xNS,\yNS) circle (\rshell);
\end{scope}

%================================================
% PROJECTED CONTRIBUTING AREA pi r_c^2
%================================================

\begin{scope}[canvas is yz plane at x=\xNS]

    \fill[
        Goldenrod!55,
        opacity=0.20
    ]
    (\yNS,\zNS) circle (\rshell);

    \draw[
        Goldenrod!80!black,
        dashed,
        line width=0.95pt
    ]
    (\yNS,\zNS) circle (\rshell);

\end{scope}

%================================================
% NEUTRON STAR
%================================================
\node[
    circle,
    shade,
    ball color=black!18,
    minimum size=1.04cm,
    inner sep=0pt,
    draw=black!48,
    line width=0.85pt
] at (NS) {};

%------------------------------------------------
% Front shell cross sections
%------------------------------------------------
\foreach \dx/\op in {
    0.35/0.20,
    0.68/0.15
}{
    \pgfmathsetmacro{\xx}{\xNS+\dx}
    \pgfmathsetmacro{\rr}{
        sqrt(\rshell*\rshell-\dx*\dx)
    }

    \begin{scope}[canvas is yz plane at x=\xx]
        \draw[
            red!78!black,
            opacity=\op,
            line width=0.48pt
        ]
        (\yNS,\zNS) circle (\rr);
    \end{scope}
}

%------------------------------------------------
% Principal great circle of the resonant shell
%------------------------------------------------
\begin{scope}[canvas is yz plane at x=\xNS]
    \draw[
        red!78!black,
        opacity=0.34,
        line width=0.66pt
    ]
    (\yNS,\zNS) circle (\rshell);
\end{scope}

%================================================
% RELATIVE MOTION
%================================================

% Motion approximately along the x direction
\draw[
    -{Latex[length=3.2mm,width=2.0mm]},
    BrickRed!90!black,
    line width=1.5pt
]
(-1.15,0.42,0.42)
--
(0.28,0.1,0.20)
node[
    midway,
    above=10pt,
    text=BrickRed!90!black,
    inner sep=1pt
]
{$v_\perp$};

%================================================
% LABELS
%================================================

\node[
    RoyalBlue!80!black,
    font=\small\bfseries,
    inner sep=1.5pt
]
at (-1.45,-1,1.30)
{cloud};

\node[
    RoyalBlue!72!black,
    font=\footnotesize
]
at (-0.95,1.92,1)
{\(\rho_{\rm cl}\simeq\rho_{\rm cl}(0)\)};

\node[
    black,
    font=\small\bfseries,
    inner sep=1pt
]
at ($(NS)+(0.03,-0.06,-0.04)$)
{NS};

\end{tikzpicture}%
}

\caption{
\(R_\star\gg r_c\): extended-cloud regime.
The neutron star and resonant-shell system moves through an
approximately homogeneous cloud.
}
\end{subfigure}

%================================================
% GENERAL CAPTION
%================================================
\caption{
Geometry of PBH-seeded axion-cloud crossings of the resonant conversion
shell.
(a) In the compact-cloud regime, \(R_\star\ll r_c\), the curvature of the
resonant shell is negligible across the cloud. Each entry or exit crossing
may therefore be described using the local tangent plane, with the geometry
defined by \(s\), \(x_\perp\), and
\(r=\sqrt{s^2+x_\perp^2}\). The thin transparent red surface represents the
full three-dimensional resonant shell surrounding the neutron star.
(b) In the extended-cloud regime, \(R_\star\gg r_c\), the neutron star and
resonant-shell system moves through a cloud whose density is approximately
constant on scales of order \(r_c\). The pale-gold disk bounded by the
dashed contour denotes the projected shell area, \(\pi r_c^2\), in the
plane perpendicular to the relative motion. At each instant, only the axion
material passing through this dashed projected area contributes to the
conversion rate. This restricted overlap produces the geometrical
suppression characteristic of the extended-cloud regime. The red solid
contour represents the resonant shell, while the gray sphere
represents the neutron star surface.
}
\label{fig:cloud_geometry_regimes}

\end{figure}
If \(s(t)=v_\perp (t-t_c)\), the photon production rate for each isolated crossing is,
\begin{equation}
\dot N_\gamma^{\rm comp}(t)
=
\frac{v_\perp}{\mu}\,
\Sigma_{\rm cl}\!\left[v_\perp(t-t_c)\right]\,
P_{a\to\gamma}^{\rm res}(r_c),
\qquad
\dot N_{\gamma,{\rm peak}}^{\rm comp}
=
\frac{v_\perp M_{\rm cloud}}{2\mu R_\star}
P_{a\to\gamma}^{\rm res}(r_c)\, .
\label{eq:Ngamma_single_crossing}
\end{equation}
The second expression follows from evaluating the column density at \(s=0\), where
\(\Sigma_{\rm cl}(0)=M_{\rm cloud}/(2R_\star)\).
In the large-cloud limit, \(R_\star\gg r_c\), the cloud density varies slowly across the resonant shell. Close to the neutron star one may approximate,
\begin{equation}
\rho_{\rm cl}(r\lesssim r_c)
\simeq
\rho_{\rm cl}(0)
=
\frac{M_{\rm cloud}}{\pi R_\star^3}\, .
\end{equation}
The mass flow through one projected resonant surface is then $\dot M_{\rm cross}
\simeq
v_\perp\,\rho_{\rm cl}(0)\,\pi r_c^2$
and therefore,
\begin{equation}
\dot N_\gamma^{\rm large}
\simeq
\frac{v_\perp\,M_{\rm cloud}}{\mu R_\star^3}
r_c^2
P_{a\to\gamma}^{\rm res}(r_c)\, .
\label{eq:Ngamma_large_cloud}
\end{equation}
Relative to the compact single-crossing peak, this corresponds to the geometrical suppression,
\begin{equation}
f_{\rm geom}
\simeq
2
\left(
\frac{r_c}{R_\star}
\right)^2,
\qquad
R_\star\gg r_c,
\qquad
P_{a\to\gamma}^{\rm res}\ll1 \, .
\label{eq:fgeom_large_cloud}
\end{equation}
The peak photon production rate is therefore,
\begin{equation}
\dot N_\gamma^{\rm peak}
\simeq
\begin{cases}
10^{50}\,{\rm s}^{-1}
\left(\dfrac{\mu}{10^{-3}\,{\rm eV}}\right)^{4/3}
\left(\dfrac{10^2}{\kappa}\right)^{5/3}
\left(\dfrac{B_0}{10^{9}\,{\rm T}}\right)^{1/3}
\left(\dfrac{\tau}{0.1\,{\rm s}}\right)^{5/3}
c_\gamma^2,
&
R_\star\ll r_c\, ,
\\[1.2em]
10^{47}\,{\rm s}^{-1}
\left(\dfrac{\mu}{10^{-3}\,{\rm eV}}\right)^4
\left(\dfrac{M_{\rm PBH}}{10^{14}\,{\rm kg}}\right)^2
\left(\dfrac{10^2}{\kappa}\right)
\left(\dfrac{B_0}{10^{9}\,{\rm T}}\right)
\left(\dfrac{\tau}{0.1\,{\rm s}}\right)
c_\gamma^2,
&
R_\star\gg r_c\, .
\end{cases}
\label{eq:Ngamma_peak_both_regimes}
\end{equation}
An important feature of Eq.~(\ref{eq:Ngamma_peak_both_regimes}) is that, for the maximal cloud-mass benchmark adopted here, the photon production rate is independent of \(f_a\). Indeed, \(M_{\rm cloud}\propto f_a^2\), while \(P_{a\to\gamma}^{\rm res}\propto g_{a\gamma\gamma}^2\propto c_\gamma^2 f_a^{-2}\), so that the explicit \(f_a\) dependence cancels. The peak luminosity and spectral flux density derived below are therefore also independent of \(f_a\) under these assumptions. For the neutron star benchmark considered here, the exterior-conversion condition
$r_c>R_{\rm NS}$ selects predominantly the extended-cloud regime. The compact-cloud regime may be reached at larger axion masses, although maintaining the resonance outside
the neutron star surface then generally requires larger surface magnetic fields. We therefore use the extended-cloud result as the reference analytical benchmark for ordinary neutron star magnetic fields. The corresponding peak luminosity in the conversion line is,
\begin{equation}
L_\gamma^{\rm peak}
=
\mu\,\dot N_\gamma^{\rm peak}
\simeq
1.6\times 10^{32}\,{\rm erg\,s^{-1}}
\left(
\frac{\mu}{10^{-3}\,{\rm eV}}
\right)^5
\left(
\frac{M_{\rm PBH}}{10^{14}\,{\rm kg}}
\right)^2
\left(
\frac{10^2}{\kappa}
\right)
\left(
\frac{B_0}{10^9\,{\rm T}}
\right)
\left(
\frac{\tau}{0.1\,{\rm s}}
\right)
c_\gamma^2 \, .
\label{eq:Lgamma_peak_extended}
\end{equation}
For the fiducial parameters, this is approximately a few per cent of the solar bolometric luminosity, but concentrated in a narrow radio or millimetre line
rather than distributed over a broad thermal spectrum. The converted photons have frequency,
\begin{equation}
\nu_{\rm conv}
=
\frac{\mu}{h}
\simeq
240\,{\rm GHz}
\left(
\frac{\mu}{10^{-3}\,{\rm eV}}
\right) \, .
\label{eq:nu_conv}
\end{equation}
Defining the fractional linewidth as
$\delta\equiv\Delta\nu/\nu_{\rm conv}$, the peak spectral flux density observed at a
distance $D$ is,
\begin{equation}
S_\nu^{\rm peak}
=
\frac{L_\gamma^{\rm peak}}
{4\pi D^2\Delta\nu}
=
\frac{h\,\dot N_\gamma^{\rm peak}}
{4\pi D^2\delta} \, .
\label{eq:Snu_from_Ndot}
\end{equation}
Using $D=10\,{\rm kpc}$ as a Galactic-centre benchmark, the extended-cloud result becomes,
\begin{equation}
S_{\nu}^{\rm peak}
\simeq
31\,{\rm Jy}
\left(
\frac{10^{-4}}{\delta}
\right)
\left(
\frac{10\,{\rm kpc}}{D}
\right)^2
\left(
\frac{10^2}{\kappa}
\right)
\left(
\frac{B_0}{10^9\,{\rm T}}
\right)
\left(
\frac{\tau}{0.1\,{\rm s}}
\right)
\left(
\frac{\mu}{10^{-3}\,{\rm eV}}
\right)^4
\left(
\frac{M_{\rm PBH}}{10^{14}\,{\rm kg}}
\right)^2
c_\gamma^2 \, .
\label{eq:Snu_peak_extended}
\end{equation}
We adopt $\delta=10^{-4}$ as a representative phenomenological benchmark. The detailed
line profile may be modified by the axion velocity distribution, plasma refraction,
gravitational ray bending and magnetospheric corotation~\cite{Battye:2021xvt}. These events correspond to bright, narrow-band transients whose duration is controlled by the spatial extent of the PBH-seeded axion cloud. In the extended-cloud regime, the duration of the full encounter is controlled by the asymptotic relative velocity, since gravitational acceleration advances the leading and trailing parts of the cloud in approximately the same way, up to tidal corrections. We therefore use,
\begin{equation}
t_{\rm sig}
\simeq
4.8\,{\rm s}
\left(
\frac{220\,{\rm km\,s^{-1}}}{v_{\rm DM}}
\right)
\left(
\frac{10^{-3}\,{\rm eV}}{\mu}
\right)^2
\left(
\frac{10^{14}\,{\rm kg}}{M_{\rm PBH}}
\right)\, .
\label{eq:signal_duration_extended}
\end{equation}
Instrumental spectral-line sensitivities are commonly quoted for a fixed reference integration
time, whereas the duration in Eq.~\eqref{eq:signal_duration_extended} varies across the
parameter space. We take $t_{\rm ref}=1\,{\rm hr}$ and define the effective integration time as,
\begin{equation}
t_{\rm eff}(\mu,B_0)
=
\min\left[
t_{\rm sig}(\mu,B_0),
t_{\rm ref}
\right]\, .
\label{eq:effective_integration_time}
\end{equation}
Thus, if the transient is shorter than one hour, the one-hour sensitivity must be degraded according to the radiometer scaling \cite{NRAO:VLA_OSS_2025A},
\begin{equation}
\sigma_{\nu,i}(\mu,B_0)
=
\sigma_{\nu,i}^{1{\rm hr}}
\left[
\nu_{\rm conv}(\mu)
\right]
\left(
\frac{t_{\rm ref}}
{t_{\rm eff}(\mu,B_0)}
\right)^{1/2}\, ,
\label{eq:transient_sensitivity_rescaling}
\end{equation}
where $i$ labels the instrument and
$\sigma_{\nu,i}^{1{\rm hr}}$ is its one-hour rms spectral-line sensitivity at the conversion frequency. The detection condition used in the numerical scan is therefore,
\begin{equation}
S_\nu^{\rm peak}(\mu,B_0)
>
5\,\sigma_{\nu,i}(\mu,B_0)\, .
\label{eq:instrument_detection_condition}
\end{equation}
For the VLA, the one-hour matched-channel rms sensitivity is calculated using \cite{NRAO:VLA_OSS_2025A},
\begin{equation}
\sigma_{\rm VLA}^{\nu,1{\rm hr}}
=
\frac{{\rm SEFD}}
{\eta_c
\sqrt{
2\,N_{\rm ant}(N_{\rm ant}-1)\,
\Delta\nu\,t_{\rm ref}
}}
\simeq
(0.1\text{--}0.5)\,{\rm mJy}
\left(
\frac{10^{-4}}{\delta}
\right)^{1/2}\left(\frac{1\,{\rm hr}}{t_{\rm ref}} \right)^{1/2}\, ,
\label{eq:VLA_line_sensitivity_reference}
\end{equation}
where we use $N_{\rm ant}=27$, a correlator efficiency
$\eta_c\simeq0.93$, and representative frequency-dependent values ${\rm SEFD}\simeq250\text{--}500\,{\rm Jy}$. The full numerical calculation interpolates the SEFD as a function of frequency and subsequently applies the transient rescaling in Eq.~\eqref{eq:transient_sensitivity_rescaling}.
In the ALMA frequency range, the spectral-line sensitivity depends strongly on the observing band, atmospheric transmission, channel width and array configuration. We therefore use representative band-dependent one-hour line sensitivities evaluated for $\delta=10^{-4}$ and apply the same duration-dependent rescaling \cite{ALMA:Cycle12PG}. Figure~\ref{fig:instrument_detection_regions} shows the regions of the $(B_0,\mu)$ plane in which the predicted peak spectral flux density satisfies the
mass-dependent $5\sigma$ condition in
Eq.~\eqref{eq:instrument_detection_condition}, for the benchmark PBH mass $M_{\rm PBH}=10^{14}\,\mathrm{kg}$. The plot additionally imposes conversion outside the neutron star surface, $r_c>R_{\rm NS}$, and the intrinsic Galactic encounter-rate requirement
$\Gamma_{\rm PBH}^{\rm Gal}\gtrsim0.1\,{\rm yr}^{-1}$. The colored regions therefore identify parameter values for which the signal is both sufficiently bright and sufficiently frequent under the benchmark assumptions. Note that we obtain detectable signals for PBH masses in the $10^{13}-10^{15}$ kg mass range and the corresponding axion mass intervals for stimulated capture, taking into account that the event luminosity (rate) increases (decreases) with the PBH mass. 

Although both the VLA and ALMA have detected transient radio and millimeter emission, existing transient searches are generally optimized for broadband and/or dispersed signals rather than narrow spectral lines lasting a few seconds. To our knowledge, no event with the combination of duration and fractional linewidth predicted here has been reported, motivating dedicated or archival searches for this distinctive signature.
\begin{figure}[t]
\centering
\includegraphics[width=0.9\textwidth]{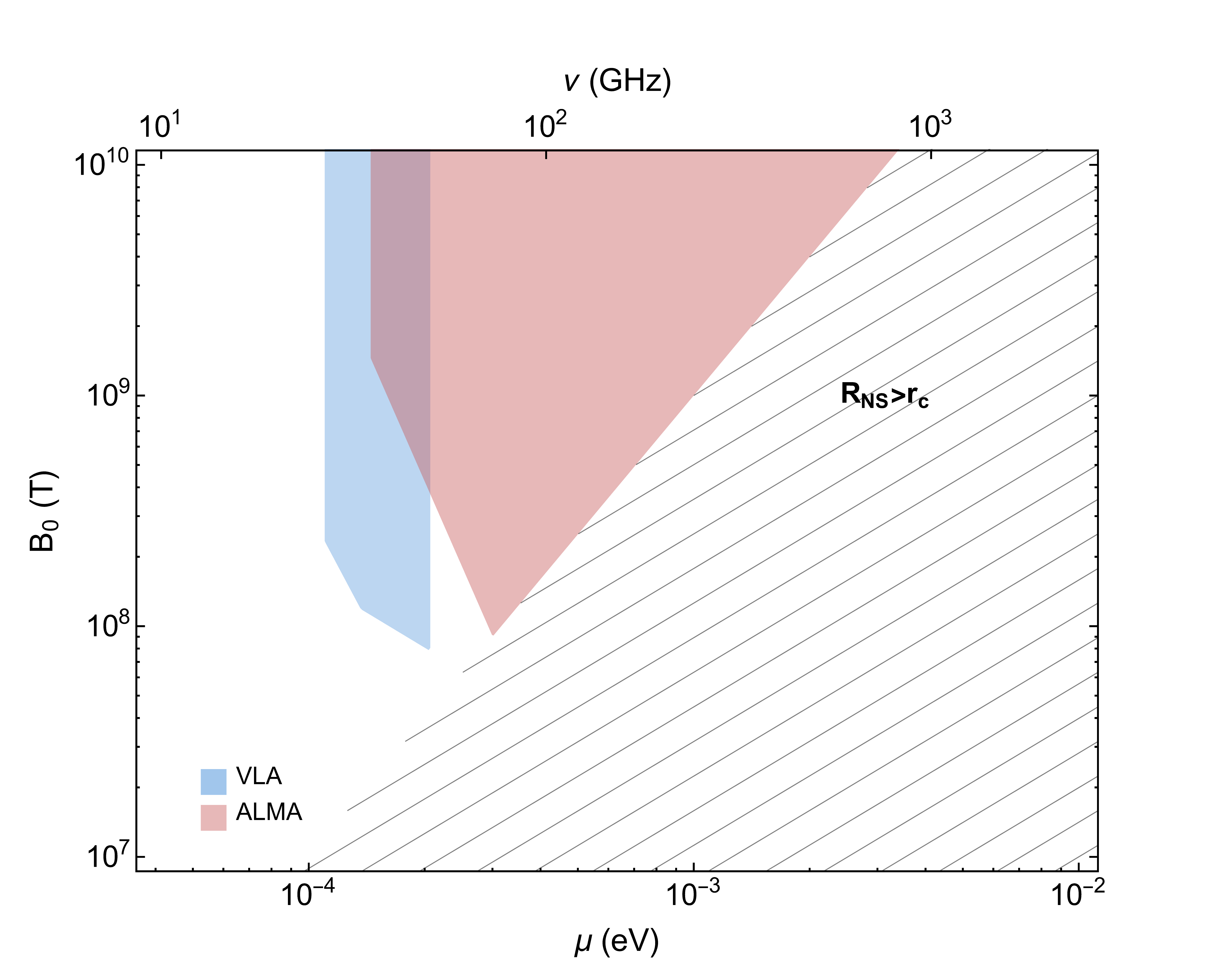}
\caption{
Detection regions in the $(B_0,\mu)$ plane for resonant axion-to-photon conversion in neutron star magnetospheres, for a PBH mass $M_{\rm PBH}=10^{14}\,\mathrm{kg}$ and $\kappa=10^2$. The blue and red shaded regions show the parameter
values for which the predicted peak spectral flux density exceeds the adopted $5\sigma$ thresholds of the VLA and ALMA, respectively
\cite{NRAO:VLA_OSS_2025A,ALMA:Cycle12PG}.
The quoted one-hour instrumental sensitivities are rescaled point by point using the effective integration time $t_{\rm eff}=\min(t_{\rm sig},1\,{\rm hr})$. The diagonally hatched region is excluded, since in this parameter range the resonant shell lies inside the neutron star, $r_c<R_{\rm NS}$, and the Galactic encounter rate falls below the adopted observational-relevance threshold, $\Gamma_{\rm PBH}^{\rm Gal}<0.1\,{\rm yr}^{-1}$.
}
\label{fig:instrument_detection_regions}
\end{figure}

% ---------------------------------------------------------
\newpage
\subsection{Microlensing and finite-size (extended-lens) effects}\label{subsec:microlensing}
% ---------------------------------------------------------

Finite-lens effects in microlensing by extended DM structures and axion stars have been studied in
Refs.~\cite{Croon:2020wpr,Fujikura:2021omw,Yin:2024xov}.
If the axion cloud carries a non-negligible fraction of the total lens mass and is spatially extended, microlensing can become an important probe. The key point is that lensing is sensitive not only to the total mass but also to its projected mass distribution. A PBH surrounded by a cloud therefore need not behave as a point lens. A simple diagnostic is the ratio between the characteristic cloud size and the Einstein radius,
\begin{equation}
\gamma \;\equiv\; \frac{R_*}{R_E}\,,
\label{eq:gamma_extendedness}
\end{equation}
where,
\begin{equation}
R_E
= \sqrt{\frac{4 M_{\rm tot}}{M^2_{\rm{Pl}}}\,\frac{D_L D_{LS}}{D_S}}
\equiv \sqrt{\frac{4 M_{\rm tot}\,D}{M^2_{\rm{Pl}}}}\,,
\qquad
M_{\rm tot}=M_{\rm{PBH}}+M_{\rm cloud}\,.
\label{eq:Einstein_radius}
\end{equation}
Here $D_L$ is the observer--lens distance, $D_S$ the observer--source distance, $D_{LS}=D_S-D_L$, and $D\equiv D_LD_{LS}/D_S$. When $\gamma\ll 1$, the lens is effectively point-like and the standard Paczy\'nski curve applies \cite{Paczynski:1985jf}. By contrast, when $\gamma\gtrsim 1$, finite-size effects become important and the magnification profile can be significantly altered. The cloud radius is estimated as,
\begin{equation}
R_* \simeq 0.65~\mathrm{mpc}\;
\left(\frac{10^{-9}}{\alpha}\right)
\left(\frac{10^{-11}\,\mathrm{eV}}{\mu}\right)\,,
\label{eq:cloud_size_estimate}
\end{equation}
showing that for light axions and $\alpha\ll1$ the cloud can be macroscopic. Thus, for sufficiently small PBH masses, and hence $R_E$, one can naturally enter the extended-lens regime.

In addition, from Eq.~\eqref{eq:Mcrit}, the axion cloud-to-PBH mass ratio is given by:
\begin{equation}
 {M_{\rm{cloud}}\over M_{\rm{PBH}}}\simeq 0.2\left({10^{-9}\over \alpha}\right)^2\left({f_a\over 10^9\ \mathrm{GeV}}\right)^2 \, ,  
\end{equation}
which further constrains the PBH and axion properties for significant extended-lens effects.

\begin{figure}[h!]
\centering
\includegraphics[width=0.75\textwidth]{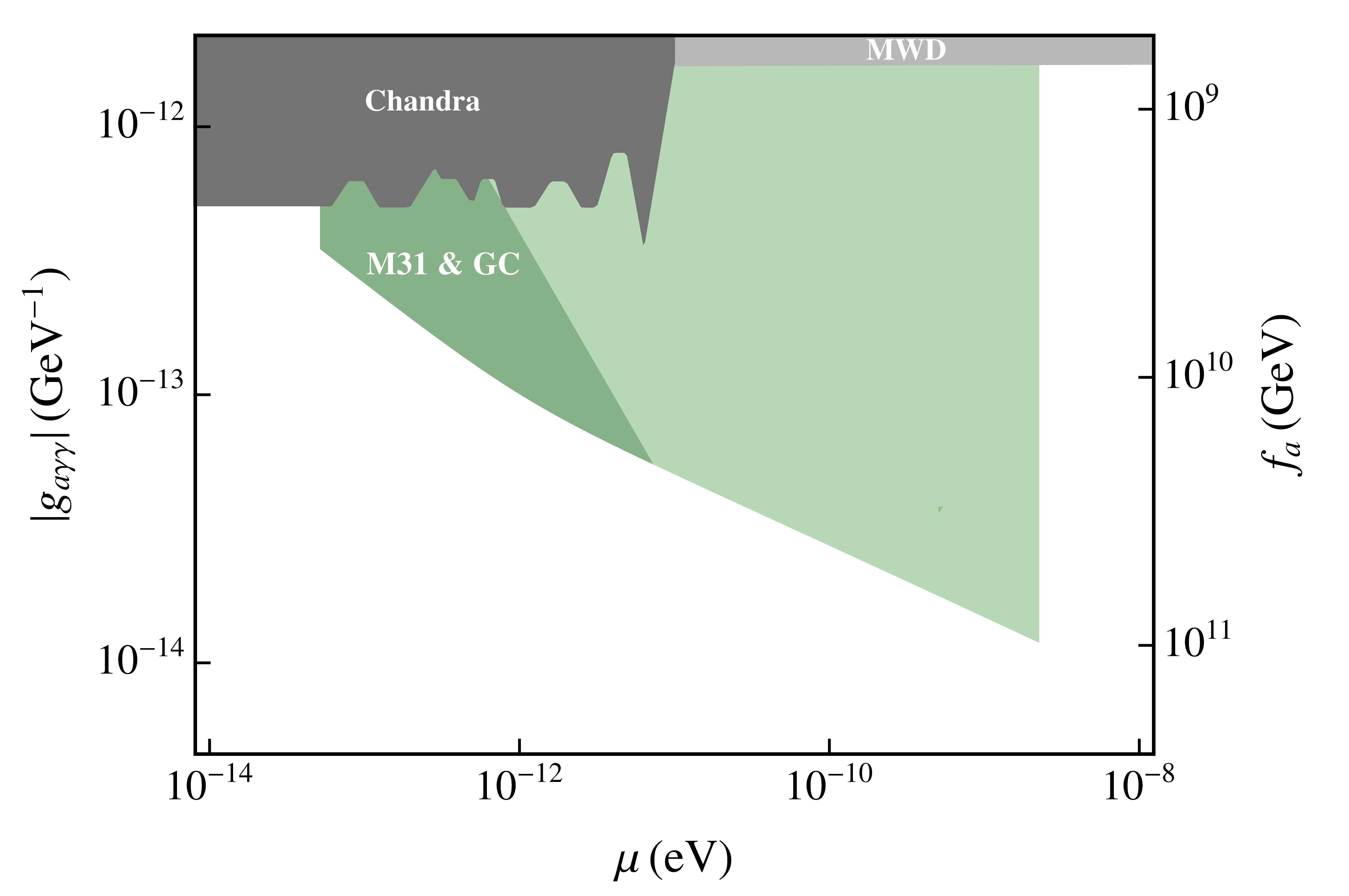}
\caption{Regions in the \((\mu,f_a)\) plane where ALP clouds may form through stimulated capture and survive until the present day around PBHs with $M_{\rm PBH}=10^{24}$ kg (light green). In the dark green sub-region, the ALP cloud mass exceeds 10\% of the PBH mass and the PBH-cloud system exhibits an extended-lens behavior  \((\gamma \equiv R_\star/R_E > 1)\), considering both sources in the Galactic Center (GC), at \(D_S=8\,\mathrm{kpc}\), and M31, at \(D_S=770\,\mathrm{kpc}\). The gray regions denote observational exclusions \cite{AxionLimits}.}
\label{fig:microlensing}
\end{figure}

To illustrate this, we consider two observational setups: OGLE-like Galactic-bulge microlensing, for which we take $D_S\simeq 8~{\rm kpc}$, consistent with the distance to the Galactic centre \cite{Gravity2019R0}, and HSC-like pixel-lensing towards M31, with $D_S\simeq 770~{\rm kpc}$ \cite{McConnachie2012}. As a fiducial and near-optimal configuration, we place the lens halfway to the source, $D_L=D_S/2$, which maximizes $D\equiv D_LD_{LS}/D_S$ (with $D\le D_S/4$). We further fix the benchmark PBH mass to $M_{\rm{PBH}}=10^{24}\,\mathrm{kg}$, motivated by recent hints in OGLE \cite{Niikura:2019kqi} and HSC--M31 \cite{Sugiyama:2026kpv} data for short-timescale candidates in this mass range. To identify the region where finite-size lensing effects become important, in Figure~\ref{fig:microlensing} we impose the extended-lens condition $\gamma>1$, alongside $M_{\rm cloud}>0.1 M_{\rm PBH}$, on top of the same ALP parameter space shown in Figure~\ref{fig:alps_region}. 

As shown in Figure~\ref{fig:microlensing}, the extended-lens regime is favored for smaller axion masses, since decreasing $\mu$ increases both the cloud size $R_\star$ and the cloud-to-PBH mass ratio, the latter also increasing with $f_a$. A similar behavior is observed for other values of $M_{\rm PBH}$ in the sub-solar mass range relevant for gravitational microlensing experiments.

\FloatBarrier
\section{Conclusion}\label{sec:conclusions}
% =========================================================

In this work we studied the stimulated capture of axion and ALP clouds around PBHs through self-scattering processes. Building on the capture mechanism of \cite{Budker:2023sex}, we identified the conditions under which an ambient axion(-like) population enters the focusing regime, where stripping is suppressed and Bose enhancement drives exponential growth, and we used a threshold-based criterion to determine when cloud growth becomes efficient compared to cosmological dilution.

For constant-mass ALPs we derived analytic estimates for the threshold epoch and for the growth efficiency at threshold, separating
the cases in which the system enters the stimulated regime during radiation domination or after matter--radiation equality. Once we
also require that axion clouds survive until the present day (against late-time stripping processes and reabsorption by the central PBH), the allowed parameter space collapses into PBH-mass-dependent islands.
A clear trend is visible in Figure~\ref{fig:alps_region}: increasing the PBH mass shifts the viable region toward smaller $\mu$ and
typically smaller $\lvert g_{a\gamma\gamma}\rvert$, leaving windows that lie below existing exclusions and only partially overlap current bounds.

We extended the same analysis to the QCD axion, emphasizing the impact of the temperature dependence of the axion mass when stimulated capture becomes efficient before the QCD crossover ($T_*>\Lambda_{\rm QCD}$), which alters both the velocity evolution and the energy density scaling prior to saturation. As shown in
Fig.~\ref{fig:qcd_region}, this leads to a split viable region spanning QCD axion masses $\mu\simeq 10^{-5}$--$10^{-1}$~eV and PBH masses roughly $M_{\rm{PBH}}\simeq 10^{12}$--$10^{19}$~kg.

Finally, we outlined potential observational implications of PBH$+$cloud systems. Axion--photon couplings can make these clouds visible
when they encounter strongly magnetized plasma environments. In particular, we considered resonant axion-to-photon conversion in
neutron star magnetospheres, where the resonance condition selects a conversion radius determined by the local plasma density. For a
Goldreich--Julian density profile enhanced by a pair multiplicity factor and a dipolar magnetic field, the conversion radius moves
outward for larger magnetic fields and pair multiplicities, and inward for heavier axions. Only configurations in which this radius
lies outside the neutron-star surface correspond to physical magnetospheric conversion.

The corresponding Landau--Zener conversion probability is typically perturbative for the benchmark couplings considered here, but
the large number of axions stored in the PBH-seeded cloud can compensate for this small probability. We estimated both the rate of PBH-neutron star encounters and the photon signal produced in a resonant-shell crossing. Including
gravitational focusing, an effective Galactic neutron star population of order $\mathcal N_{\rm NS}\sim10^9$ gives an encounter rate of order one event per decade for the used benchmark values.

The resulting electromagnetic signal can be very bright in flux density under optimistic assumptions about the cloud mass. In the extended-cloud regime, where the cloud size is larger than the resonant shell radius the peak duration is controlled by the cloud size and by the local DM asymptotic  velocity. For the maximal cloud-mass benchmark used here, extended-regime peak flux densities can reach
$S_\nu^{\rm peak}\sim31\,\mathrm{Jy}$ at Galactic-centre distances for a fractional linewidth
$\delta\sim10^{-4}$. Notably, for the maximal cloud-mass benchmark adopted here, the $f_a^2$ scaling of the cloud mass exactly compensates the $f_a^{-2}$ dependence of the resonant conversion probability, so that the predicted peak flux density is independent of $f_a$, while retaining its dependence on the model-dependent coefficient $c_\gamma$.
The predicted line flux densities can remain well above nominal VLA and ALMA sensitivities in parts of the parameter space, provided the event-rate and
exterior-conversion cuts are satisfied. These estimates demonstrate that neutron star magnetospheres can act
as efficient electromagnetic transducers for otherwise dark PBH-seeded axion clouds.

In addition, axion clouds surrounding PBHs need not behave as point lenses. Whenever the cloud size becomes comparable to or larger than the Einstein radius, $\gamma \equiv R_\star/R_E \gtrsim 1$, finite-size effects can potentially modify the microlensing signal. For the benchmark mass $M_{\rm{PBH}}=10^{24}\,\mathrm{kg}$, Figure \ref{fig:microlensing} shows that this condition is satisfied in a substantial portion of the surviving ALP parameter space, additionally requiring $M_{\rm cloud}>0.1M_{\rm PBH}$. The extended-lens regime is reached most readily for smaller axion masses, which produce larger and heavier clouds. Although the extended-lens condition is generally more restrictive for M31-like configurations due to their larger Einstein radius, the parametric region obtained for Galactic-bulge and M31-like sources overlap completely within the parameter space shown. This demonstrates that PBH$+$cloud systems can naturally lead to potentially observable departures from the standard point-lens Paczy\'nski description, motivating a dedicated study of their microlensing phenomenology in future work.

Overall, our results show that stimulated capture around PBHs can populate long-lived axion and ALP clouds in viable regions of parameter space with potentially detectable signals. Stimulated capture is thus a compelling alternative mechanism to the more widely studied superradiant instability as a mechanism for axion cloud formation. The two mechanisms operate in complementary parametric regimes and have different requirements for efficient cloud growth. Superradiance, on the one hand, generates axion/ALP clouds irrespective of their contribution to DM, but requires PBH to be spinning sufficiently fast and very suppressed self-interactions; stimulated capture, on the other hand, is efficient for Schwarzschild BHs but only if axions are a significant fraction of DM and exhibit non-negligible self-interactions. Moreover, superradiance populates an excited, axially symmetric bound state, while stimulated capture populates dominantly the spherically symmetric ground state.

Although in this work we have focused on PBHs heavier than $10^{12}$ kg and, hence, lifetimes exceeding the age of the Universe, it would be interesting to investigate cloud formation for lighter PBHs and the axion clouds after PBH evaporation. According to the numerical analysis in \cite{Neves:2025kxp}, under certain adiabaticity conditions clouds may survive as self-gravitating bound states, and we plan in future work to investigate this possibility in more detail in the context of stimulated capture.

We thus hope that our results motivate further investigations of PBH-seeded axion-cloud formation, its observational signatures, and its relevance
for fundamental particle physics and cosmology.

\section*{Acknowledgements}
We thank Ricardo Ferreira and Rodrigo Vicente for useful discussions on topics related to this work. NPB is supported by the FCT doctoral fellowship PRT/BD/154617/2022. This work was partially supported by national funds by FCT - Funda\c{c}\~ao para a Ci\^encia e Tecnologia, I.P., through the research project with DOI identifier 10.54499/UID/04564/2025.

\appendix

\section{Description of axion cloud growth}
\label{app:dense_growth}

In this appendix we give a more detailed explanation of the dynamics of axion cloud formation around PBHs. At early times, when $\xi\ll1$, stimulated capture is subdominant and
the cloud evolves approximately according to the rates used in Section \ref{sec:axions_general}:
\begin{equation}
 \dot M_{\rm cloud}\simeq C-\Gamma_2M_{\rm cloud},\qquad
 C\simeq\frac{36\rho_a^3\alpha^4}{v_a^9f_a^4\mu^6},\qquad
 \Gamma_2\simeq\frac{2\rho_a^2\alpha^4}{v_a^6f_a^4\mu^3}\, .
 \label{eq:app_capture_rates}
\end{equation}
The corresponding equilibrium mass is
\begin{equation}
 M_{\rm cloud}^{\rm eq}=\frac{C}{\Gamma_2}
 \simeq\frac{18\rho_a}{v_a^3\mu^3}\, .
 \label{eq:app_equilibrium_mass}
\end{equation}
For a constant-mass ALP, after field oscillations begin at $H\sim \mu$, we have $\rho_a\propto a^{-3}$ and $v_a\propto a^{-1}$. Thus, within the $\xi\ll1$ approximation,
both $C$ and $\Gamma_2$ are constant, and a cloud initially empty at
$t_i$ evolves as
\begin{equation}
 M_{\rm cloud}(t)\simeq M_{\rm cloud}^{\rm eq}
 \left[1-e^{-\Gamma_2(t-t_i)}\right]\, ,
 \label{eq:app_initial_growth}
\end{equation}
until $t\sim t_*$ at which the PBH-cloud system enters the focusing regime where stimulated capture becomes the dominant process. At this stage, the axion cloud has grown to a mass:

\begin{equation}
 M_{\rm seed}\equiv M_{\rm cloud}(t_*)
 =\frac{C}{\Gamma_2}\left[1-e^{-\Gamma_2(t_*-t_i)}\right]\, .
 \label{eq:app_seed_general}
\end{equation}
For constant-mass ALPs, we have $H(t_i)\simeq\mu\simeq (2t_i)^{-1}$ in the radiation era and
$v_a(t_i)\simeq1$.
Using $v_a(t_*)=2\pi\alpha$, stimulated capture becomes dominant at:
\begin{equation}
 t_*\simeq
 \begin{cases}
 t_R, & t_R<t_{\rm eq}\\[2pt]
 t_R\left(t_{\rm eq}/t_R\right)^{1/4}, &  \quad t_R>t_{\rm eq}
 \end{cases}
 ~,\qquad t_R\equiv\frac{1}{8\pi^2\mu\alpha^2}\, .
 \label{eq:app_threshold_time}
\end{equation}
Here we have used $a\propto t^{1/2}$ in radiation domination and
$a\propto t^{2/3}$ after matter-radiation equality, matched at $t_{\rm eq}$.
Since the cosmological axions correspond to an oscillating condensate in the conventional misalignment picture,
$\rho_a(t_i)\simeq f_a^2\mu^2$, with $v_a(t_i)\simeq 1$.
Since $\rho_a\propto a^{-3}$ and $v_a\propto a^{-1}$,
we have $\Gamma_2\simeq 2\mu\alpha^4$.
This implies that:
\begin{equation}
\Gamma_2 t_*\simeq
\frac{\alpha^2}{4\pi^2}
\begin{cases}
1, & t_*<t_{\rm eq},\\[4pt]
(8\pi^2\mu\alpha^2t_{\rm eq})^{1/4},
& t_*>t_{\rm eq}\, .
\end{cases}
\end{equation}
Since $\alpha\ll 1$ we find
$\Gamma_2t_*\ll 1$ in both regimes. Thus, stripping
can be neglected during the initial capture stage,
and the cloud mass grows approximately linearly,
\begin{equation}
M_{\rm seed}\equiv M_{\rm cloud}(t_*)
\simeq C(t_*-t_i)\simeq Ct_*\,.
\label{eq:app_seed_linear}
\end{equation}
Substituting the expressions for $t_*$, we obtain
\begin{equation}
M_{\rm seed}\simeq
\frac{9f_a^2\alpha^2}{2\pi^2\mu}
\begin{cases}
1,
& t_*<t_{\rm eq},\\[5pt]
(8\pi^2\mu\alpha^2t_{\rm eq})^{1/4},
& t_*>t_{\rm eq}\, .
\end{cases}
\label{eq:app_seed_alp}
\end{equation}
For $t>t_*$, stimulated capture becomes dominant
and the cloud grows exponentially,
\begin{equation}
\dot M_{\rm cloud}\simeq\Gamma_1M_{\rm cloud}\,.
\end{equation}
For constant $\mu$, we have
$\Gamma_1\propto\rho_a^2/v_a^2\propto a^{-4}$.
Integrating, we obtain for the final cloud mass at $t\gg t_*$:
\begin{equation}
M_{\rm cloud}^{\rm final}\simeq
M_{\rm seed}
\exp\left[
\beta\left.\frac{\Gamma_1}{H}\right|_{t_*}
\right],
\qquad
\beta\simeq
\begin{cases}
1/2, & t_*<t_{\rm eq},\\
2/5, & t_*>t_{\rm eq}\, .
\end{cases}
\label{eq:app_stimulated_growth}
\end{equation}
Thus, the axion cloud reaches its maximum possible mass,
$M_{\rm cloud}^{\rm max}\simeq
8\pi f_a^2/(\alpha\mu)$, provided that:
\begin{equation}
\left.\frac{\Gamma_1}{H}\right|_{t_*}
\gtrsim
\beta^{-1}
\ln\left(
\frac{M_{\rm cloud}^{\rm max}}{M_{\rm seed}}
\right)
\simeq
\beta^{-1}
\begin{cases}
\displaystyle
\ln\left(\frac{16\pi}{\alpha^3}\right),
& t_*<t_{\rm eq},\\[10pt]
\displaystyle
\ln\left[
\frac{16\pi}
{\alpha^3(8\pi^2\mu\alpha^2t_{\rm eq})^{1/4}}
\right],
& t_*>t_{\rm eq}\, .
\end{cases}
\label{eq:app_growth_condition}
\end{equation}
For the range of $\alpha$ considered
in this work, the logarithmic factors in
Eq.~\eqref{eq:app_growth_condition} vary approximately
between $40$ and $70$. Given its weak logarithmic dependence on the model parameters, we adopt the
simplified growth criterion
$\left.\Gamma_1/H\right|_{t_*} \gtrsim 60\,\beta^{-1}$
in our analysis.

For the QCD axion, the evolution is somewhat different given the initial temperature dependence of the axion mass, which is particularly relevant for scenarios where stimulated capture becomes significant before the QCD crossover, $T_*>\Lambda_{\rm QCD}$. In this case, we find that the axion cloud's occupation number (rather than its mass) grows linearly in the $\xi<1$ regime (given that $C/\mu=const.$) where it could at most reach an equilibrium value $N_{\rm cloud}^{\rm eq}\simeq 18\rho_a/(v_a^3\mu^4)$.

Once stimulated capture becomes dominant at
$t_*<t_{\rm QCD}$, the corresponding rate
$\Gamma_1$ is, in this case, independent of the scale factor, given
that $\rho_a\propto a$, $v_a\propto a^{-5}$ and
$\mu\propto a^4$. The ground state's occupation
number thus grows exponentially such that:
\begin{equation}
N_{\rm cloud}(t_{\rm QCD})
\simeq N_{\rm cloud}(t_*)
e^{\Gamma_1(t_{\rm QCD}-t_*)}~.
\end{equation}
Afterwards, cloud growth proceeds as for a constant-mass
ALP, so that the final cloud mass can be written as:
\begin{equation}
M_{\rm cloud}^{\rm final}\simeq
\mu_0 N_{\rm cloud}(t_*)
\exp\left[
\frac{1}{2}
\left.\frac{\Gamma_1}{H}\right|_{t_*}
\left(\frac{2t_{\rm QCD}}{t_*}-1\right)
\right]~.
\end{equation}
Thus, the exponential growth of a QCD axion cloud is
more pronounced in the case $T_*>\Lambda_{\rm QCD}$
than in the opposite regime, where growth proceeds as
for a constant-mass ALP. However, we find that $T_*$ is only at most a few times larger than $\Lambda_{\rm QCD}$, so that the growth enhancement (relative to a constant-mass ALP) is not too large. 
Therefore, we (conservatively) also impose the simplified growth criterion
$\left.\Gamma_1/H\right|_{t_*} \gtrsim 60\,\beta^{-1}$ in the case of QCD axion clouds.

\newpage
\bibliography{mybibfile}

\end{document}